\documentclass[fleqn,usenatbib]{mnras}

\usepackage{newtxtext,newtxmath}

\usepackage[T1]{fontenc}
\usepackage{ae,aecompl}
\usepackage{booktabs}

\usepackage{enumitem}%
\usepackage{calc}
\usepackage{graphicx}	
\usepackage{amsmath}	
\usepackage{amssymb}	
\usepackage[dvipsnames]{xcolor}
\usepackage{xcolor}
\usepackage{makecell}

\newcommand{\kms}{km\,s$^{-1}$}	

\definecolor{mypink}{rgb}{0.958, 0.188, 0.478}

\newcommand{\StarNet}{\texttt{StarNet} }

\title[Machine learning for optical stellar spectra]{Assessing the performance of LTE and NLTE synthetic stellar spectra in a machine learning framework}

\author[Bialek et al.]{
Spencer Bialek,$^{1}$\thanks{E-mail: sbialek@uvic.ca.  
The code used in this analysis is available at  \href{https://github.com/Spiffical/StarNet}{https://github.com/Spiffical/StarNet}
}
S\'ebastien Fabbro,$^{1,2}$
Kim A. Venn,$^{1}$
Nripesh Kumar,$^{1,3}$ 
\and
Teaghan O'Briain,$^{1}$
Kwang Moo Yi$^{4}$
\\$\;$
\\
$^{1}$Department of Physics and Astronomy, University of Victoria, Victoria, BC, V8W 3P2, Canada\\
$^{2}$National Research Council Herzberg Astronomy \& Astrophysics, 4071 West Saanich Road, Victoria, BC, Canada \\
$^{3}$Department of Computer Science, National Institute of Technology, Tiruchirappalli, India \\
$^{4}$Department of Computer Science, University of Victoria, Victoria, BC, V8P 5C2, Canada
}

\date{Accepted XXX. Received YYY; in original form ZZZ}

\pubyear{2020}

\begin{document}
\label{firstpage}
\pagerange{\pageref{firstpage}--\pageref{lastpage}}
\maketitle



\begin{abstract}
In the current era of stellar spectroscopic surveys, synthetic spectral libraries are the basis for the derivation of stellar parameters and chemical abundances. 
In this paper, we compare the stellar parameters determined using five popular synthetic spectral grids (INTRIGOSS, FERRE, AMBRE, PHOENIX, and MPIA/1DNLTE) with our convolutional neural network (CNN, \texttt{StarNet}).  
The stellar parameters are determined for six physical properties (effective temperature, surface gravity, metallicity, [$\alpha$/Fe], radial velocity, and rotational velocity) given the spectral resolution, signal-to-noise, and wavelength range of optical FLAMES-UVES spectra from the Gaia-ESO Survey. 
Both CNN modelling and epistemic uncertainties are incorporated through training an ensemble of networks. 
\texttt{StarNet} training was also adapted to mitigate differences between the synthetic grids and observed spectra by augmenting with realistic observational signatures (i.e. resolution matching, wavelength sampling, Gaussian noise, zeroing flux values, rotational and radial velocities, continuum removal, and masking telluric regions). 
Using the FLAMES-UVES spectra for FGK type dwarfs and giants as a test set, we quantify the accuracy and precision of the stellar label predictions from \texttt{StarNet}.
We find excellent results over a wide range of parameters when \StarNet is trained on the MPIA/1DNLTE synthetic grid, and acceptable results over smaller parameter ranges when trained on the 1DLTE grids.
These tests also show that our CNN pipeline is highly adaptable to multiple simulation grids.
\end{abstract}


\section{Introduction}



Astronomy has entered an era of spectroscopic surveys.  Over the past two decades, the remarkably successful Sloan Digital Sky Survey \citep[SDSS][]{York2000} provided the first spectroscopic survey of a large number of stars \citep[c.f.,][and other sources]{Yanny2009}, soon followed by the RAdial Velocity Experiment \citep[RAVE, c.f.,][]{Steinmetz2006, Steinmetz2020} survey of nearly a half million stars, and the Large Sky Area Multi-Object Fibre Spectroscopic Telescope \citep[LAMOST,][]{cui2012}, which has collected spectra for $\sim$1 million stars \citep[e.g.,][]{zhang2019}.  
These spectroscopic surveys were carried out with low resolution optical spectra (R$\sim$2000, SDSS and LAMOST) or medium resolution spectra in a narrow wavelength range (R=7500, 841-880\,nm, RAVE).  More recently, higher resolution spectroscopic surveys have been initiated with enormous success in the determination of stellar parameters and chemical abundances, including nearly half a million stars in the SDSS Apache Point Observatory Galactic Evolution Experiment survey \citep[APOGEE, R$\sim$22,500, 1.51-1.70\,$\mu$m, c.f.,][]{holtzman2018} and over 350,000 stars in the Galactic Archaeology with HERMES survey \citep[GALAH, R$\sim$40,000, 400-700\,nm, c.f.,][]{buder2018}.  Deeper optical spectroscopic surveys will soon begin in 2020 at the 4-metre telescopes, including the WHT Enhanced Area Velocity Explorer (WEAVE) survey \citep[e.g.,][]{Dalton2018} and the European Southern Observatory  4-metre Multi-Object Spectroscopic Telescope (4MOST) survey \citep[e.g.,][]{dejong2019, 4MOST2019}, both providing high and low resolution observing modes.  Also, the 8-metre Subaru Telescope will initiate a very deep Galactic Archaeology survey using their optical and near-IR 3-arm Prime Focus Spectrograph \citep[PFS, R$\sim$5000, 380-1260\,nm, e.g.,][]{PFS2018}. 


To prepare for this era of large data sets, methods to consistently and efficiently analyse stellar spectra are being explored, particularly with sophisticated data analysis algorithms, e.g., ``The Cannon" \citep{Ness2015, buder2018}, ``The Payne" \citep{ting2019, Xiang2019}, and ``MATISSE" \citep{Matisse2006, Kordopatis2013}.  
In addition to these methods, we have developed our own convolutional neural network, \StarNet
\citep{fabbro2018}.
\StarNet reproduces the stellar parameters of benchmark stars and predicted the stellar parameters for the entire APOGEE spectral data set within minutes. Furthermore, \StarNet 
could be trained either from data with a priori known stellar labels (data-driven mode) or from a synthetic spectral grid (synthetic mode). \citet{leung2019} improved on the {\it data-driven version} of \StarNet by modifying the neural network architecture to track individual abundances, to train on missing or noisy stellar labels, and to estimate prediction uncertainties. 


Although there have been comparisons made of synthetic spectra libraries \citep[e.g.,][]{martins2019testing}, currently lacking is a comparison of the uncertainties and the issues related to their application to real data when used with machine learning tools. In this paper, we examine the impacts of training \StarNet with a variety of publicly available high resolution, optical synthetic stellar grids. The synthetic grids include
 INTRIGOSS \citep{franchini2018gaia}, AMBRE \citep{delaverny2012}, PHOENIX \citep{husser2013}, FERRE \citep{allendeprieto2018}, and a grid of spectra that includes NLTE corrections for H, O, Mg, Si, Ca, Ti, Cr, Mn, Fe, and Co (hereafter named `MPIA' since the spectral synthesis online tool is hosted at the Max Planck Institute for Astronomy, \citealt{NLTE_MPIA}). 
 These grids of synthetic spectra have been generated using independent model atmospheres and radiative transfer codes (all 1DLTE), with a range of atomic and molecular opacities required to describe the stellar photosphere.  

 %
 %

In this analysis, and for the first time, several different \textit{optical} synthetic spectral grids are used to train a convolutional neural network, which \StarNet is ideally suited for.  Upgrades to \StarNet are described in Section 2, including a new deep ensembling method to provide estimates of the uncertainties in the stellar labels. 
We also describe our efforts to pre-process and ``augment'' any set of synthetic grids (to a common resolution, wavelength sampling, and continuum normalization scheme, and by including observational signatures) to produce realistic training sets and overcome the synthetic gaps. In Section 3, the synthetic grids studied in this paper are introduced and compared.  Three grids are chosen to train and test \StarNet in Section 4: (1) the semi-empirical INTRIGOSS 1DLTE spectral grid, (2) the FERRE 1DLTE grid, and (3) the 1DNLTE MPIA grid.  The other spectral grids are used for testing, validation, and comparisons of the predicted labels and uncertainties.  In Section 5, the three trained \StarNet models are applied to the FLAMES-UVES spectra from the Gaia-ESO Survey to test the performance of each model on observational spectra.  Our results, and caveats, of training a neural network on synthetic spectra are discussed in Section 6, including future plans to further develop \StarNet for the quick analysis of spectra from the new Gemini Observatory GHOST spectrograph.

\section{Methods \label{section:methods}}









\subsection{Analysis with neural networks}

Only a brief description of neural networks is provided here to establish the terminology used in this paper. See \cite{fabbro2018} for a more complete description of \StarNet and our machine learning methodology.  

Fundamentally, a neural network (NN) is a function which transforms an input to a desired output. The function is composed of many parameters, arranged in layers, which form a highly non-linear combination of the input features, allowing for complex mappings to be represented accurately. 
\StarNet is a \textit{convolutional} NN, in which a series of learned filters, followed by a series of learned inter-connected nodes, transform a stellar spectrum to a prediction of associated stellar parameters.


To ensure the NN does not over- or under-fit the data, the full data set is typically split into a training, validation, and test set. The training set is used to directly influence the parameters of the NN, and the validation set is used to periodically check the performance of the NN on a separate data set. Both of these sets are utilized during the training of the NN, in which data is iteratively sent through the NN, the parameters of the NN are nudged in a direction which minimizes the output of the \textit{loss function} (for regression problems, the loss is typically the residual between the prediction and expected output). In this study, the final model is updated throughout training as the iteration which performs best on the validation set. Since both the training and validation sets influence the final trained NN, the test set is used to quantify the final performance for an independent data set. 

A potential alternative to a NN discriminative method would be a physically motivated forward modelling approach. Within a Bayesian framework, built-in uncertainty quantification is offered \citep[e.g., ][]{schonrich2014bayesian, schneider2017bonnsai}. Delivering full Bayesian posteriors over stellar parameters and abundances can be very resource intensive for survey-size data sets, even with modern Markov Chain Monte Carlo speed ups. Given our practical goals of obtaining quick and robust uncertainties for a given survey, we pursue the same CNN approach as in \cite{fabbro2018} with its trade-offs, and also allow the NN to learn uncertainty predictions. 

For a training set of 90,000 spectra, each with $\sim$40,000 flux values, the training time for \StarNet rarely exceeds 30 minutes using a single Tesla V100 GPU. 
With a final trained model, predictions for a set of thousands of spectra can be completed in seconds,  allowing on-line interactive analysis.

\subsection{Modifications to \StarNet}

\subsubsection{Uncertainty Predictions}

To derive predictive uncertainties we have adapted the method of \textit{deep ensembling}, in which an ensemble of probabilistic NNs with different initialization are trained, as outlined in \cite{lakshminarayanan2017simple}. 
%
%
Each NN can predict a probability density function (PDF) for the physical parameters of interest. In this study, the PDF is assumed to be Gaussian to simplify the comparisons of the spectral grids and to allow for efficient analysis of millions of spectra -- this assumption can be generalized to more complex and asymmetric PDFs, e.g., a Gaussian mixture \citep{disanto2018mdn}. The mean and standard deviation of each predicted Gaussian PDF, after ensemble averaging, is associated to the predictive uncertainty of each stellar parameter.  Good statistical coverage has been shown for this simple deep ensembling method, including the epistemic uncertainties accounting for the NN modelling and for out-of-distribution samples \citep{ovadia2019can}. It is relatively simple to implement, and required only a few small changes to our original \StarNet architecture, as described here:
\begin{enumerate}[labelindent=8pt,labelwidth=\widthof{},itemindent=0em,leftmargin=!]

   \item The NN of input spectra $\mathbf{x}$ and target predictions $y$ outputs a parametric PDF $p_\theta(y|\mathbf{x})$ capturing aleatoric uncertainties. In our case, the last layer of the NN predicts both the mean $\mu_\theta(\mathbf{x})$ and a learned variance $\sigma_\theta^2(\mathbf{x})$ of a Gaussian distribution.
   
   \item A proper scoring rule is used for a training loss function. For our regression use case, the score is the negative log-likelihood for a normal distribution:
   \begin{equation}
    \label{eq:ensemble_loss}
    -\log{p_{\theta}(y|\mathbf{x})} = \frac{\log{\sigma^2_\theta(\mathbf{x})}}{2} + \frac{(y - \mu_\theta(\mathbf{x}))^2}{2\sigma^2_\theta(\mathbf{x})}.
   \end{equation}
   
   \item An ensemble of $M$ NNs (typically 5-7) are trained with different random initialization. At test time, all $M$ NN predictions are combined such that 
   \begin{equation}
    \label{eq:deepprob}
        p(y|\mathbf{x})=\frac{1}{M}\sum_{m=1}^M p_{\theta_m}(y|\mathbf{x}).  
    \end{equation}
   The final prediction, $\mu_*(\mathbf{x})$, and final variance, $\sigma_*^2(\mathbf{x})$, can be obtained by combining the outputs from each model as $\mu_*(\mathbf{x})$ is given by the average of the predicted means of each NN, and the final variance is determined via the following equation:
   \begin{equation}
    \label{eq:deepuncertainty}
    \sigma_*^2(\mathbf{x}) = \frac{1}{M} \sum_{m=1}^{M}\left(\sigma^2_{\theta_m}(\mathbf{x})  + \mu^2_{\theta_m}(\mathbf{x})\right) - \mu^2_*(\mathbf{x})
   \end{equation}
   This ensembling recipe allows the inclusion of epistemic uncertainties in the final prediction.

\end{enumerate}

The method of deep ensembling is a significant upgrade from the original \StarNet architecture because of its ability to quantify how closely the spectra in a test set resemble the spectra used to train the model.  The estimated uncertainty not only covers the uncertainty due to the finite sample training size, but also some of the out-of-distribution uncertainties. In contrast to the Monte-Carlo dropout method for uncertainty predictions, it does not perturb the network architecture as much, and has been shown to be well calibrated \citep{ovadia2019can}. Furthermore each model can be trained efficiently in an embarrassingly parallel mode.


\subsection{Augmenting and pre-processing the data \label{section:augmenting}}

Observed spectra typically have different shapes and profiles compared to synthetic spectra due to instrumental impacts and other signature effects. 
Special care is required to ensure both sets of spectra are standardized to minimize this \textit{synthetic gap}. 

Several steps to address and reduce the synthetic gap are involved, including (i) pre-processing the spectra (i.e., matching the resolution and sampling of the spectra to a common wavelength scale and removing the continuum) and (ii) augmenting the spectra data sets (e.g., adding Gaussian noise, rotational and radial velocities, masking telluric regions, and zeroing flux values to mimic bad pixels). Augmenting data is a popular method used in machine learning experiments, serving to increase the robustness of the NN to variations that exist in reality and to increase the size of a training dataset.  Synthetic spectral grids typically contain several thousand templates; however, more data is usually required for NN training.  

Our process for creating an augmented synthetic dataset required actions on randomly selected spectra in the original dataset.
Each spectrum from the original set was therefore chosen several times and given different augmentations.  To prepare for our application to the Gaia-ESO survey (see Section 5), 
the following modifications (in order) were applied to the synthetic spectra:


\begin{enumerate}[labelindent=8pt,labelwidth=\widthof{},itemindent=0em,leftmargin=!]
  \item \textit{Resolution matching}: spectra were convolved to a resolution of the UVES spectra (R $\sim$ 47,000)
  \item \textit{Rotational velocity}: randomly chosen with the constraint 0\,\textless{}\,$v_{\textrm{rot}}$\,\textless{}\,50\,km/s
  \item \textit{Radial velocity}: randomly chosen with the constraint |$v_{\textrm{rad}}$|\,\textless{}\,150\,km/s which covers the Gaia-ESO range
  \item \textit{Sampling matching}: the wavelength grid was re-sampled onto the UVES wavelength grid
  \item \textit{Additive noise}: Gaussian noise was added with the constraint $\sigma$\,\textless{}\,7\% flux value, corresponding to S/N (per pixel)\,\textgreater{}\,14
  \item \textit{Continuum removal}: using the method described in Appendix~\ref{appendix:continuum}
  \item \textit{Data imputation}: random samples of the synthetic flux values were set to zero, with a maximum of 10\% of the spectrum 
  \item \textit{Tellurics masking}: known telluric lines\footnote{Telluric lines from the Keck-MAKEE pipeline, available online at https://tinyurl.com/y4f5flpx} are given a value of zero.
\end{enumerate}

All of the modifications up to and including the continuum removal [(i)-(vi) above], were pre-computed in parallel before training.
The last two items were applied to the generated spectra on-the-fly during training.


\section{Synthetic Spectral Grids
\label{sec:datasets}}

\begin{table*}
\caption{The parameter space and sampling of the synthetic spectra grids used in this study.  *Note that the 1DNLTE MPIA spectra must be generated using an online tool, and chemical abundances and micro/macro-turbulent velocities can be varied at will; we applied [$\alpha$/Fe] and $v_{\textrm{micro}}$ limits to match the INTRIGOSS grid.
\label{tab:all_synthetic_grids} }
\begin{tabular}{@{}lccccccccccccccc@{}}
\toprule
          & \multicolumn{3}{c}{$T_{\textrm{eff}}$ (K)} & \multicolumn{3}{c}{log$g$ (dex)} & \multicolumn{3}{c}{{[}Fe/H{]} (dex)} & \multicolumn{3}{c}{{[}$\alpha$/Fe{]} (dex)} & \multicolumn{3}{c}{$v_{\textrm{micro}}$ (km/s)} \\ \cmidrule(l){2-4} \cmidrule(l){5-7} \cmidrule(l){8-10} \cmidrule(l){11-13}
          \cmidrule(l){14-16}
          & Min.    & Max.     & Step    & Min.     & Max.     & Step     & Min.        & Max.       & Step      & Min.         & Max.        & Step       & Min.      & Max.      & Step      \\ \midrule
INTRIGOSS & 3750    & 7000     & 250     & 0.5      & 5.0      & 0.5      & -1.0        & 0.5        & 0.25      & -0.25        & 0.5         & 0.25       & 1         & 2         & 1         \\ \midrule
FERRE     & 3500    & 6000     & 500     & 0        & 5.0      & 1        & -5.0        & 0.5        & 0.5       & \multicolumn{3}{c}{{0.5 at [Fe/H] $\leq$ -1.5}}        & 1.5        & 1.5        & --        \\
          & 5500    & 8000     & 500     & 1.0      & 5.0      & 1        & -5.0        & 0.5        & 0.5       & \multicolumn{3}{c}{\makecell{0.0 at [Fe/H] $\geq$ 0 \\ linear in between}}        & 1.5        & 1.5        & --        \\ \midrule
AMBRE     & 2500    & 8000     & 250     & -0.5     & 5.5      & 0.5      & -5.0        & 1.0        & 0.25      & -0.4         & 0.4         & 0.2        & 1        & 2        & 1        \\ \midrule
PHOENIX   & 2300    & 7000     & 100     & 0        & 6.0      & 0.5      & \makecell{-4.0 \\ -2.0} & \makecell{-2.0 \\ 1.0} & \makecell{1.0 \\ 0.5} & -0.2         & 1.2         & 0.2        & 0        & 4        & f($T_{\textrm{eff}}$)        \\
          & 7000    & 15000    & 200     & 0        & 6.0      & 0.5      & \makecell{-4.0 \\ -2.0} & \makecell{-2.0 \\ 1.0} & \makecell{1.0 \\ 0.5} & -0.2           & 1.2          & 0.2         & 0        & 4        & f($T_{\textrm{eff}}$)        \\ \midrule
MPIA*   & 4600    & 8800     & 200     & 1.0        & 5.0      & 0.2      & -4.8  & 0.9 & 0.3 & -0.25         & 0.5         & 0.25        & 1        & 2        & 1       \\ 
         \bottomrule
\end{tabular}
\end{table*}

There are numerous synthetic spectral grids available in the literature \citep[e.f., see][]{martins2017grids}, each differing in their spectral parameter and wavelength ranges typically to reflect the goals in a specific scientific survey (e.g., SDSS, LAMOST, APOGEE). Each has also been generated with different assumptions for the model atmospheres, radiative transfer codes, and atomic and molecular data. For example, an assumption made in most spectrum synthesis codes is that of local thermodynamic equilibirium (LTE) in the stellar atmosphere, but some codes do correct for non-LTE (NLTE) effects to produce more realistic absorption profiles in the synthesized spectra. All of these differences can have significant impacts on the synthetic spectra, making direct comparisons between synthetic grids, and observed spectra, challenging and 
inconsistent. 

To train a machine learning algorithm, it is necessary to carefully consider which grid of synthetic spectra is best to use for a particular spectroscopic survey and/or science case.  

\subsection{The synthetic grids used in this study}

The synthetic spectra examined in this analysis include the high (spectral) resolution grids INTRIGOSS, AMBRE, FERRE, and PHOENIX, and we have generated a new 1DNLTE grid using a synthetic spectral generator hosted at the Max Planck Institute for Astronomy (MPIA). All spectra in the grids were obtained with absolute fluxes, and then continuum normalized as outlined in Appendix \ref{appendix:continuum}.
 
The parameter space covered by the grids is summarized in Table~\ref{tab:all_synthetic_grids}, and a brief description of each grid follows: 

\begin{enumerate}[labelindent=8pt,labelwidth=\widthof{},itemindent=0em,leftmargin=!]

\item \textbf{INTRIGOSS:} this is a set of high resolution synthetic spectra generated from ATLAS12 model atmospheres and SPECTRUM v2.76f radiative transfer code, and tailored for the analysis of F, G, and K type stars in the Gaia-ESO survey \cite{franchini2018gaia}. The INTRIGOSS synthetic spectra allow the stellar parameters $T_{\textrm{eff}}$, log$g$, [Fe/H], [$\alpha$/M], and v$_{\textrm{micro}}$ to vary within relatively small ranges (see Table \ref{tab:all_synthetic_grids}) and span the wavelength range 483-540\,nm only. This wavelength range is a subset of the entire wavelength range of the FLAMES/UVES spectra (480-680 nm, in three settings), but it contains important features, such as H$\beta$, the Mgb lines, and numerous metal lines.  INTRIGOSS is unique in that it was fine-tuned with astrophysical gf-values through comparisons with a very high S/N solar spectrum and the UVES-U580 spectra of five cool giants (all with [Fe/H] $\sim -1$).
In some cases, the line list was (semi-empirically) modified to match the observed spectra without identifying the source of the feature.  

\item \textbf{FERRE:} a medium and high resolution collection of synthetic spectral grids generated with ATLAS9 model atmospheres and the ASSET radiative transfer code \citep{koesterke2008center}, and prepared to be used with the FERRE optimization code \citep{prieto2006spectroscopic}.
It covers a wide wavelength range (120-6500\,nm) and large parameter space (3500 $\leq T_{\textrm{eff}} \leq$ 30,000 K, 0 $\leq$ log$g \leq$ 5, -5 $\leq$ [Fe/H] $\leq$ 1. 
The grids chosen for this study had coarse parameter sampling because they were the only options available at high resolution (R\,$\ge$\,100,000), though different parameter sampling and resolution combinations (e.g. the finer grids include [$\alpha$/Fe] as an independent dimension) are available \citep{allendeprieto2018}.
These spectra reproduce the main absorption features from the UV to the near IR for B to early-M type stars, and have been used recently in the spectral analyses of stars in the SDSS (APOGEE, MaNGA, eBOSS) and Pristine surveys \citep[e.g., see][]{leung2019, Aguado2019SDSS, aguado2019}.
The full FERRE grid is split into 5 sub-grids with increasing ranges in temperature; only the first two are used in this study.

\begin{figure*}
	\includegraphics[width=2.0\columnwidth]{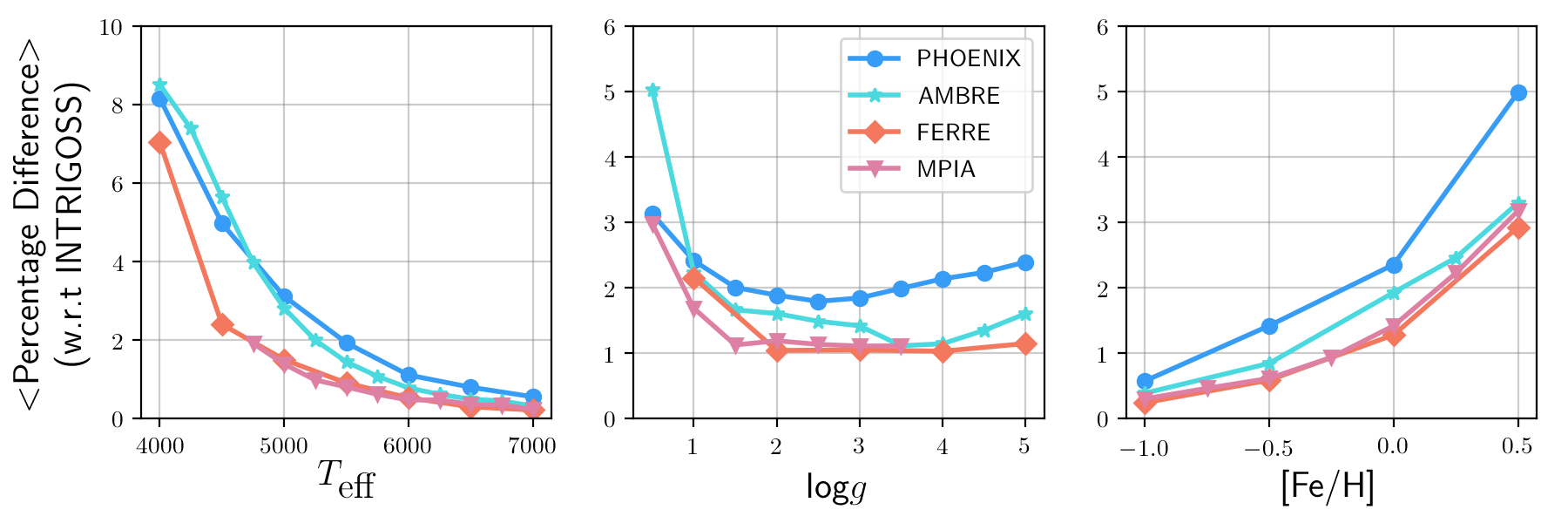}
    \caption{Comparison of synthetic spectra from the five grids examined in this paper.  For each INTRIGOSS spectrum, synthetic spectra from the PHOENIX, AMBRE, FERRE, and MPIA grids were collected with the same range of stellar parameters.  The percentage difference was calculated per spectrum, and median differences determined as a function of temperature, surface gravity, and metallicity. This plot is very similar to the comparisons made by \citealt{franchini2018gaia} in their analysis of their INTRIGOSS spectra (their Figure 7). }
    \label{fig:percent_diff_synthetic_spectra}
\end{figure*}

\item \textbf{AMBRE:} this is a high resolution (R\,$>$\,150,000) optical spectral grid (300-1200\,nm) generated from MARCS model atmospheres and the LTE Turbospectrum code for F, G, K, and M type stars.  Four stellar parameters are varied over a relatively large extent (2500\,$\leq$\,$T_{\textrm{eff}}$\,$\leq$\,8000\,K, -0.5\,$\leq$\,log$g$\,$\leq$\,5.5, -5\,$\leq$\,[Fe/H]\,$\leq$\,1, -0.4\,$\leq$\,[$\alpha$/Fe]\,$\leq$\,0.4). This grid was generated several years ago after adopting the atomic data in the VALD3 database \citep{delaverny2012}, although a more recent version also includes a range in neutron-capture elements [s-,r-/Fe] \citep{Guiglion2018}. It has also been used to predict stellar parameters in the Gaia-ESO UVES survey \citep[e.g.,][]{worley2016ambre}.

\item \textbf{PHOENIX:} this spectral grid was generated using PHOENIX model atmospheres and radiative transfer code \citep{husser2013}. All model atmospheres were calculated assuming LTE and spherical geometry.  NLTE was included for a few special lines (e.g., Li I, Na I, K I, Ca I, and Ca II).  The synthetic spectra were generated with very high resolution (R\,$>$\,100,000) spanning a wide wavelength range from the UV to mid-IR (50-5000\,nm).  The grid covers a large parameter space (2300\,$\leq$\,$T_{\textrm{eff}}$\,$\leq$\,12,000\,K, 0\,$\leq$\,log$g$\,$\leq$\,6, -4\,$\leq$\,[M/H]\,$\leq$\,1, -0.2\,$\leq$\,[$\alpha$/M]\,$\leq$\,1.2), and 
was used to analyse MUSE integral field spectra of stars in the metal-poor globular cluster NGC~6397 \citep{husser2016}.  More recently, it has also been used in a machine learning application, i.e., in the analysis of lower resolution LAMOST spectra \citep{Wang2019}.

\item \textbf{MPIA:} We have generated an NLTE synthetic spectral grid using a new online spectrum synthesis tool\footnote{MPIA spectra were generated with the \href{http://nlte.mpia.de}{MPIA online tool}} using MAFAGS-OS model atmospheres \citep{NLTE_MPIA, grupp2004a, grupp2004b} and NLTE atomic data \citep{2007A&A...461..261M, 2008A&A...492..823B,2010A&A...522A...9B,2010MNRAS.401.1334B,2011MNRAS.413.2184B, 2012ApJ...751..156B,2012MNRAS.427...27B,2013ApJ...764..115B,2013AstL...39..126S,2015ApJ...804..113B,2017ApJ...847...15B}.  MAFAGS-OS has been designed for A, F, and G stars, and the spectrum synthesis includes departures from LTE in the line formation of several species (H\,I, O\,I, Mg\,I, Si\,I, Ca\,I/II, Ti\,I/II, Cr\,I, Mn\,I, Fe\,I, and Co\,I), which are expected to more accurately model the majority of the absorption features. This helps to reduce the synthetic gap (see Section \ref{subsection:comparisons_of_synthetic_grids}), particularly for metal-poor stars where NLTE effects can be large \citep{jofre2015gaia, kovalev2019non, mashonkina2019}.
We used the online tool to batch synthesize spectra with a specified resolution, wavelength range, set of stellar parameters (limited to 4600\,K\,$\leq T_{\textrm{eff}} \leq$\,8800\,K, 1.0\,$\leq$ log$g \leq$\,5.0, -4.8\,$\leq$ [Fe/H] $\leq$\,0.9), and dispersion in [$\alpha$/Fe] ratios, and we have made it publicly available\footnote{Raw MPIA spectra created in this work are available \href{https://zenodo.org/record/3972378}{here}}. 
\end{enumerate}



\subsection{Comparisons of synthetic grids}
\label{subsection:comparisons_of_synthetic_grids}

\begin{figure*}
	\includegraphics[width=2.0\columnwidth]{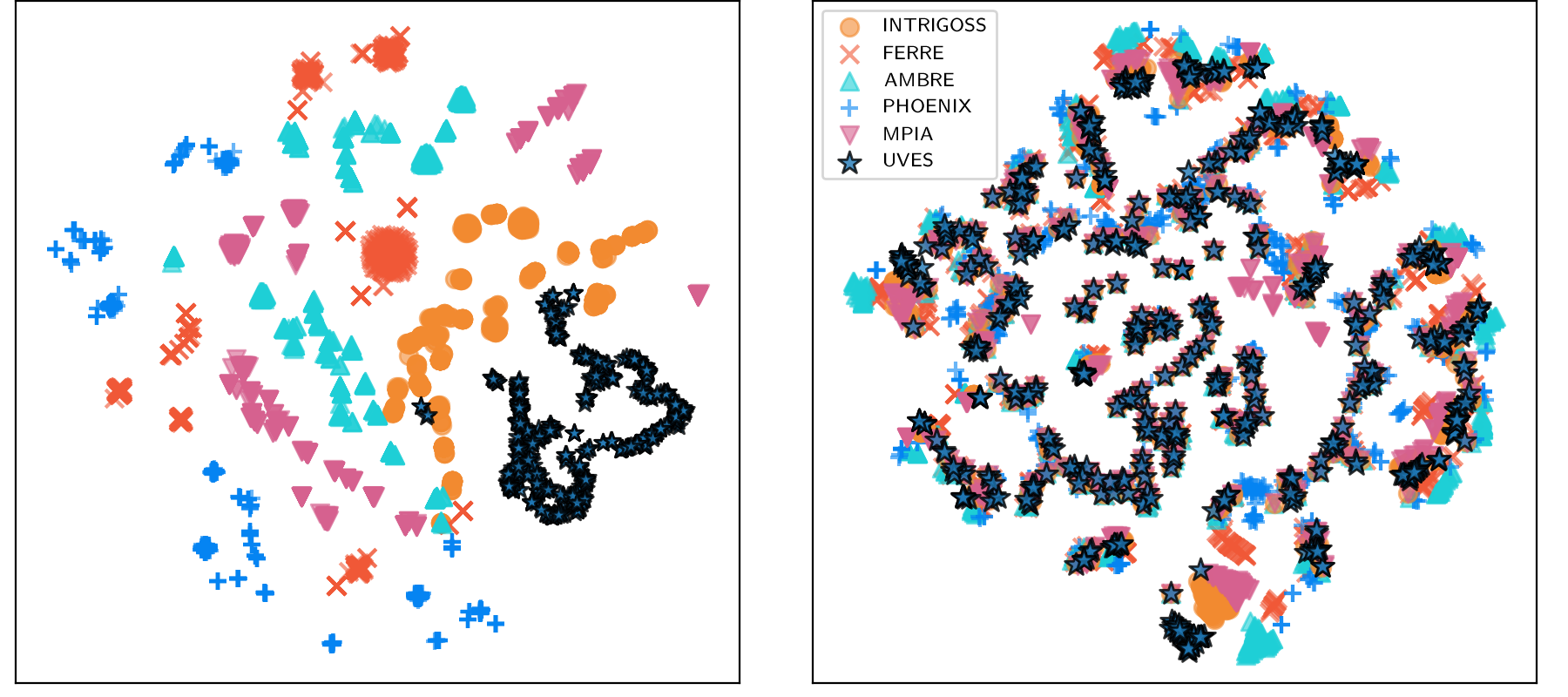}
    \caption{t-SNE plots to visualize any synthetic gaps between the five synthetic spectral grids used in this analysis (INTRIGOSS, FERRE, PHOENIX, AMBRE, and MPIA) and the observed Gaia-ESO UVES spectra (grey points). Left panel is the raw synthetic and observational data, showing the largest synthetic gaps.  Right panel shows pre-processed and augmented synthetic spectra, where the synthetic gaps are mostly overcome. For each UVES spectrum, the synthetic spectrum from each grid was collected with the closest matching parameters to the associated GES iDR4 values. Note: in this t-SNE dimensionality reduction, the 2 dimensions are not physical and units are not interpretable, but distance between points quantify the similarity between datasets. 
    \label{fig:tsne_synthetic_grids}}
\end{figure*}





Grid sampling was not a primary focus in this study, however it is certainly worth further investigation for its effects on the quality of interpolation between grid points. 
A more quantitative study on the effect of sampling strategies and adaptive simulation-based studies is delayed for future investigation.

To perform a comparison of the 1DLTE synthetic spectral grids, INTRIGOSS was selected as the baseline.  For each INTRIGOSS spectrum, spectra with matching stellar parameters from each grid were selected (and if none were found, the INTRIGOSS spectrum was skipped). The residuals of the flux values of each spectrum with respect to the INTRIGOSS spectrum were calculated and converted to a percentage difference. The average percentage difference was then determined in bins of temperature, surface gravity, and metallicity. As shown in Figure \ref{fig:percent_diff_synthetic_spectra}, the differences in the spectra are more pronounced at lower temperatures and higher metallicities, i.e., in the grid regions that would be the most sensitive to atomic and molecular transitions.  Furthermore, the FERRE 1DLTE spectra are closely matched to the semi-empirical INTRIGOSS grid, over the widest range in stellar parameters. In the space of normalized fluxes, the FERRE grid is not very different from the 1DNLTE MPIA grid.  The PHOENIX spectral grid shows the largest deviations from all of the other grids. 

To qualitatively assess how closely the synthetic spectral grids match the Gaia-ESO FLAMES-UVES spectra 
(see Section \ref{section:testing_intrigoss_on_uves} for a full explanation of this data set), spectra from each grid, with stellar parameters most closely matched to the UVES spectra stellar parameters  (retrieved from the Gaia-ESO Survey Data Release 4), were collected and pre-processed to have the same resolution, wavelength sampling, and continuum normalization as the UVES spectra. A t-SNE\footnote{T-distributed Stochastic Neighbor Embedding (t-SNE) is a nonlinear dimensionality reduction technique well-suited for embedding high-dimensional data for visualization in a low-dimensional space of two or three dimensions.  It is often used to visualize high-level representations learned by a NN. The similarity of each sample is encoded in the overlap, or clustering, present in this low-dimensional space} 
test was then carried out to compare the closest matching spectra from each grid to each UVES spectrum.
As seen in the left panel of Figure \ref{fig:tsne_synthetic_grids}, the t-SNE reveals a distinct difference between the observed and synthetic spectra; the \textit{synthetic gap}.  However, when the data is augmented with simulated noise prior to the removal of their continuum (as described in Section~\ref{section:augmenting}), then the synthetic gap is significantly narrowed and the augmented synthetic spectra occupy the same compressed low-dimensional space as the observed FLAMES-UVES spectra, as seen in the right panel of Figure \ref{fig:tsne_synthetic_grids}.


\section{Training and testing StarNet on synthetic spectra}
\label{section:training_intrigoss}

Following standard machine learning methods for mitigating under- and over-fitting, the 1DLTE INTRIGOSS, FERRE, AMBRE, and PHOENIX spectral grids, and the 1DNLTE MPIA grid, were split into \textit{reference} and \textit{test} sets (an 80/20 split). These datasets were pre-processed and augmented (as described in Section \ref{section:augmenting}) to create datasets several times their size: 100,000 spectra for the INTRIGOSS reference set and 200,000 each for the other grids, each with test sets of 10,000 spectra. 
The reference sets were then further split into \textit{training} and \textit{validation} sets (a 90/10 split). These augmented sets of spectra were used to both train \StarNet and to analyze the results of the training procedures.

For a better comparison, each training set of spectra was constrained to the same parameter space as INTRIGOSS, except for metallicity which was allowed to extend to [Fe/H] $\ge$ -3. When trained on the MPIA grid (or INTRIGOSS, FERRE, AMBRE, or PHOENIX), then the resulting CNN model is referred to as \texttt{"StarNet-MPIA"} (or  \texttt{"StarNet-INTRIGOSS"}, \texttt{"StarNet-FERRE"}, \texttt{"StarNet-AMBRE"}, or \texttt{"StarNet-PHOENIX"}, respectively). 
 
\subsection{Method-dependent systematic biases}
\label{section:methoddependentbiases}

As a first application to examine and minimize systematic uncertainties, \StarNet was trained on each of the augmented spectral grids separately.   
Since the spectral properties (stellar parameters and continua) of each synthetic spectrum are known \textit{a priori}, then we can examine and mitigate errors or degeneracies.

In this Section, we present our CNN models and parameter comparisons from only three (of the five) spectral grids: (1) INTRIGOSS, because it has been semi-empirically calibrated specifically for the wavelength regions of the Gaia-ESO survey (the spectral region we are highlighting in this paper, and will ultimately test with comparison to VLT UVES data); (2) MPIA, because our preliminary results show it provides excellent results compared to the Gaia-ESO survey benchmark stars \citep{heiter2015gaia, jofre2014gaia, jofre2018gaia}, and has a physical basis for its line formation theory that extends over a wide range of parameters and wavelength regions; and (3) FERRE, because it is commonly used in spectroscopic surveys and our results suggest that the spectra resemble the MPIA and INTRIGOSS grids more than AMBRE and PHOENIX  (see Figure \ref{fig:percent_diff_synthetic_spectra}). Examination of the \StarNet model results based on the AMBRE and PHOENIX synthetic grids are presented in Appendix B.





\begin{figure*}
	\includegraphics[width=2.0\columnwidth]{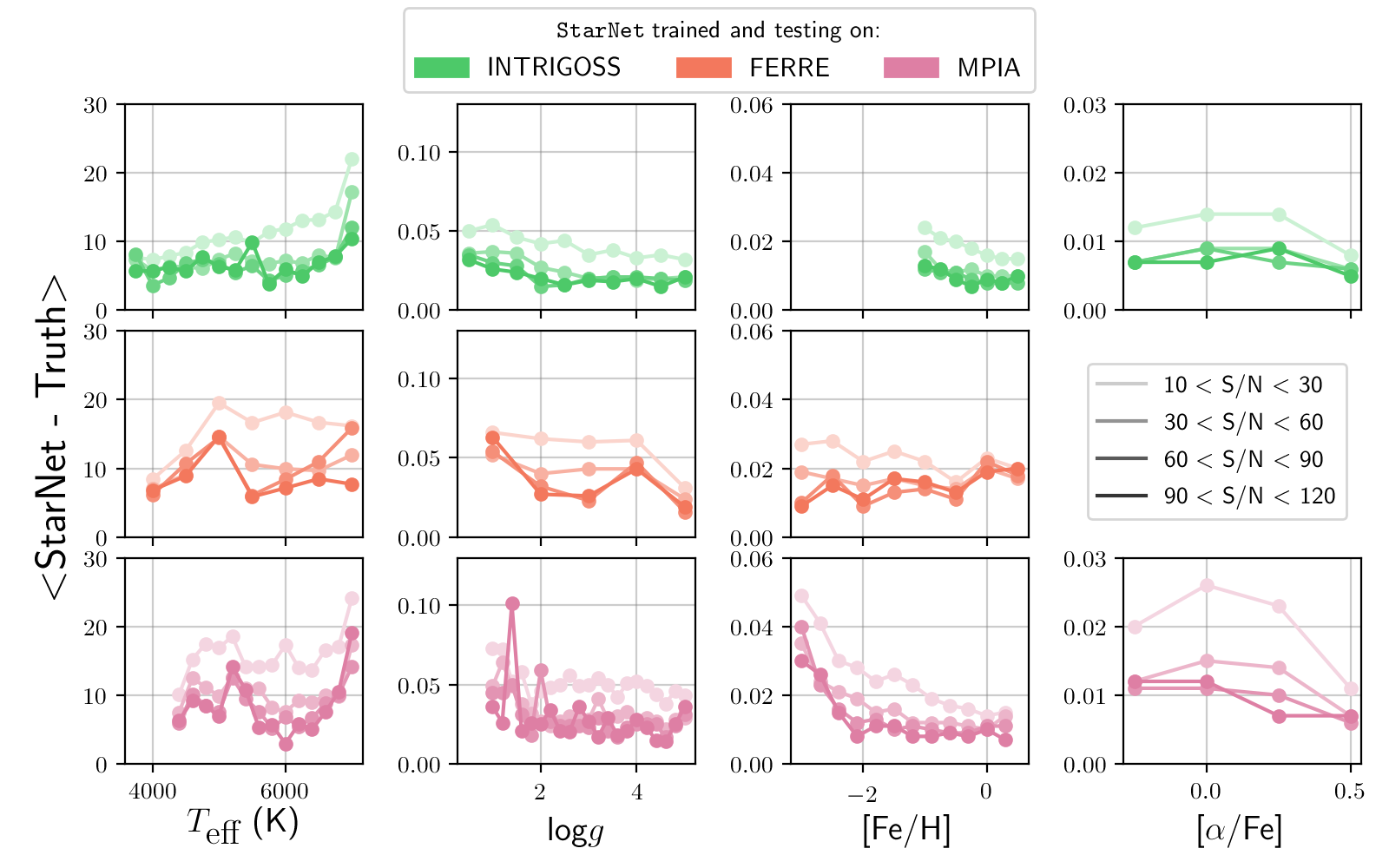}
    \caption{\StarNet was separately trained on INTRIGOSS, FERRE, and MPIA spectra. The median absolute residuals of predictions for respective test sets of 10,000 augmented spectra each, split into four signal-to-noise bins, were derived. The systematics of \StarNet as a function of both the parameter ranges and their dependence on noise are shown.}
    \label{fig:starnet_systematics}
\end{figure*}

Test sets of 10,000 augmented INTRIGOSS, FERRE, and MPIA spectra, with the same parameter ranges, were held out during training and used to identify potential systematic errors in each trained \StarNet model. Figure \ref{fig:starnet_systematics} shows the median prediction errors of each model across the four main stellar parameters (excluding [$\alpha$/Fe] for \texttt{StarNet-FERRE}, which is not an independent dimension in the FERRE grid) as a function of the signal-to-noise (S/N) ratio.  The results of these tests set the minimum uncertainties in the predictions from each \texttt{StarNet} model, and identify parameters where the uncertainties are inherently larger.

\begin{enumerate}[labelindent=8pt,labelwidth=\widthof{},itemindent=0em,leftmargin=!]

\item \textbf{Temperature, $T_{\textrm{eff}}$;} discrepancies are larger at higher temperatures in all three cases, especially at low S/N values.  This is expected due to the smaller number of spectral features available in the hotter spectra, and the increasing degeneracy in the determination of temperature and gravity in warmer stars.

\item \textbf{Surface gravity, log$g$;} the accuracy in log g are fairly constant over the parameter range tested in all cases.

\item \textbf{Metallicity, [Fe/H];}  For stars with [Fe/H] $ \ge -1.0$, the metallicity recovery is sub-percent accurate in all cases and over a wide range in S/N values.  The FERRE model appears to maintain small uncertainties to very low metallicities, near [Fe/H] $= -3.0$, whereas the uncertainties from the MPIA 1DNLTE grid imply increasing errors with lower metallicity. The same trend on 1DNLTE spectra was observed by \cite{kovalev2019non} in their analysis of FGK stars in the Gaia-ESO survey.

\item \textbf{Chemical abundances, [$\alpha$/Fe];} This abundance ratio appears to be accurate in the INTRIGOSS and MPIA models, within the parameter ranges of each grid (recall, INTRIGOSS only applies to models with [Fe/H] $\ge -1.0$, and FERRE does not treat [$\alpha$/Fe] as an independent variable).  The errors increase significantly when the S/N of the spectra decreases below $\sim30$ in the parameter and wavelength ranges tested in both models.

\item \textbf{Rotational and radial velocities, $v_{\textrm{rot}}$ and $v_{\textrm{rad}}$;} Both velocities are recovered with small uncertainties across all S/N values when trained on all three synthetic grids ($\le$0.5\,km/s in $v_{\textrm{rot}}$, and $\le$0.18\,km/s in $v_{\textrm{rad}}$). 

\end{enumerate}




\subsection{Testing \texttt{StarNet-MPIA}
with the other Synthetic Grids
\label{subsection:testing_uncertainties_synthetic} }

\begin{figure*}
	\includegraphics[width=2.0\columnwidth]{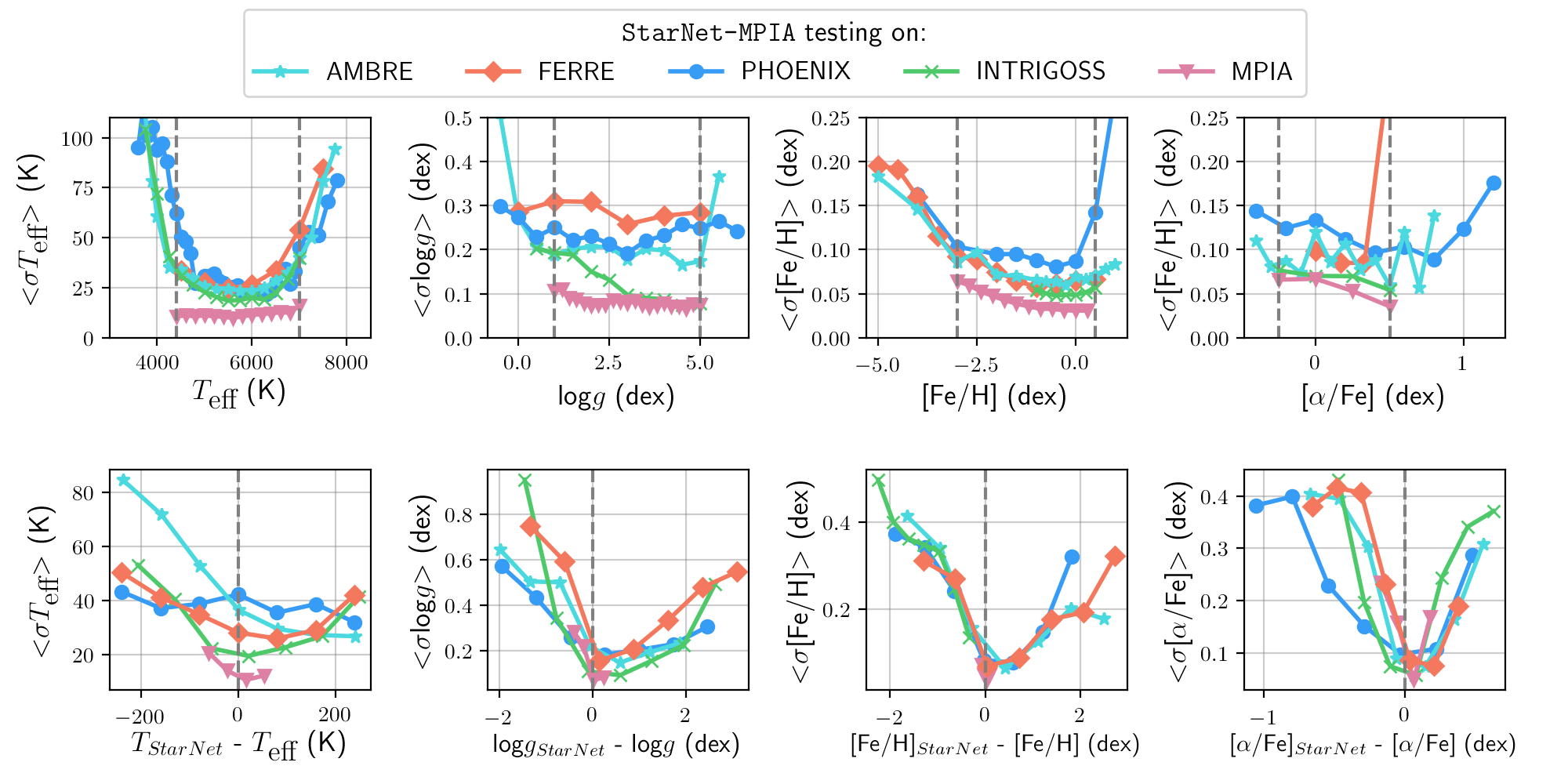}
    \caption{The predicted uncertainties of \texttt{StarNet-MPIA} for the four main stellar parameters. For each test set of 10,000 augmented  INTRIGOSS, AMBRE, FERRE, PHOENIX, and MPIA spectra, the binned median uncertainties were calculated. Top row shows the predicted uncertainties vs. the ground truth values, and the second row shows the predicted uncertainties vs. the difference between \StarNet parameter predictions and the ground truth values.  In general, the uncertainties grow when predicting outside of the parameter range trained on (shown as vertical dashed lines in the first row), and when discrepancies between predictions and truth are large (second row).}
    \label{fig:errors_all_synthetic_intriparams}. 
\end{figure*}

%

%
In this Section, we predict stellar parameters and uncertainties for the augmented test sets from INTRIGOSS, AMBRE, FERRE, and PHOENIX, while setting \texttt{StarNet-MPIA} as the reference training set.
The discrepancies in stellar parameter estimates and predicted uncertainties are summarized in Figure \ref{fig:errors_all_synthetic_intriparams}, including  for the MPIA test set (discussed in the previous section) for completeness. 

The predicted uncertainties from the 1DLTE grids increase relative to the uncertainties from the MPIA training set at lower temperatures, lower surface gravities, and higher metallicities (i.e. where the synthetic grids were previously shown to deviate the mos, see Figure \ref{fig:percent_diff_synthetic_spectra}). 
As expected, the uncertainties tend to increase when predicting outside of the parameter ranges used for training, as well as when the predictions become more discrepant from their true values. This fact confirms the predicted \StarNet uncertainties do include epistemic uncertainties.

The differences in temperature and metallicity are similar for each of the 1DLTE stellar grids, however the uncertainties in surface gravity predictions vary significantly. In the top row of Figure \ref{fig:errors_all_synthetic_intriparams}, the uncertainties in gravity from the FERRE spectra appear to be $\sim$3 times larger than from the MPIA test set, and the most offset from all of the other 1DLTE grids. In contrast, the uncertainties in gravity from the INTRIGOSS spectra are closely matched to those of the MPIA test set, especially at higher gravities. Indeed, we note that all of the predicted uncertainties from the INTRIGOSS grid are closest to those of the MPIA grid, suggesting these grids of spectra are the most similar. This result highlights the success of the NLTE corrections -- derived from first principles and thus widely applicable -- in matching the limited ad hoc corrections of INTRIGOSS that were based on matching synthetic absorption features to observed features. 

\section{Gaia-ESO FLAMES-UVES test set}
\label{section:testing_intrigoss_on_uves}

In addition to testing \StarNet with the synthetic spectra generated from a variety of radiative transfer and model atmosphere codes, we also evaluate our pipeline with \textit{observed optical} spectra from the Gaia-ESO public spectroscopic survey \citep[GES,][]{Gilmore2012}.
This is a large optical survey aiming to explore all components of the Milky Way and is complementary to Gaia. Along with the observed spectral database, an official Gaia-ESO Survey Internal Data Release (GES iDR) is available, containing stellar spectra and stellar parameters derived as the weighted average of the results from a set of working groups (each using different methods).   The fourth data release (GES iDR4) is used in this study as a comparison for \StarNet predictions \citep{Pancino2017}.

The GES was carried out using the FLAMES spectrograph at the VLT \citep{pasquini2002installation} which has two branches: the GIRAFFE instrument was used to obtain high-quality medium-resolution spectra for 10$^5$ stars, and the UVES instrument collected high-resolution (R $\sim$ 47,000) spectra for $\sim$5,000 stars.
A dataset of 2,308 FLAMES-UVES spectra is used in our analysis, spanning field and cluster stars from the bulge, halo, thick disc and thin disc of the Milky Way.

The GES also includes a set of 34 \textit{benchmark} spectra of well-known bright dwarfs, sub-giants, and giants \citep{blancocuaresma2014} which can be used as a reference set, and is available online\footnote{\url{ftp://obsftp.unige.ch/pub/sblancoc/Gaia_Benchmark_Stars_Library/}}. The benchmark stars' stellar parameters $T_{\textrm{eff}}$ and log$g$ were determined independently from spectroscopic indicators, i.e., using angular diameter measurements and bolometric fluxes \citep{heiter2015gaia}, while their [Fe/H] and [$\alpha$/Fe] parameters were determined via spectroscopic measurements with NLTE corrections \citep{jofre2015gaia}.

In this section, we apply \texttt{StarNet-MPIA}, \texttt{StarNet-FERRE}, and \texttt{StarNet-INTRIGOSS} to the Gaia-ESO spectral database and compare the results to the GES-iDR4 stellar parameters.  \texttt{StarNet-AMBRE} and \texttt{StarNet-PHOENIX} evaluations are presented in Appendix B.

\subsection{StarNet predictions for the GES benchmark stars}

\begin{table*}
\caption{A comparison of stellar parameter results from \StarNet trained on the INTRIGOSS, FERRE, and MPIA augmented grids and applied to GES benchmark stars. MRD = metal rich dwarfs, MRG = metal rich giants, and MP = metal poor stars.  The average quadratic differences (see text) between the StarNet predictions and the GES benchmark star parameters (for those stars only within the parameter ranges trained on) are shown.}
\label{tab:threemodels_benchmarks}
\hspace*{-0.0cm}\begin{tabular}{@{}lcccccccccccc@{}}
\toprule
 & \multicolumn{4}{c}{MRD (7 stars)} & \multicolumn{4}{c}{MRG (3 stars)} & \multicolumn{4}{c}{MP (7 stars)} \\
 \cmidrule(lr){2-5}\cmidrule(lr){6-9}\cmidrule(lr){10-13}
 & $\overline{\Delta}$$T_\textrm{eff}$ & $\overline{\Delta}$log$g$ & $\overline{\Delta}${[}Fe/H{]} & $\overline{\Delta}${[}$\alpha$/Fe{]} & $\overline{\Delta}$$T_\textrm{eff}$ & $\overline{\Delta}$log$g$ & $\overline{\Delta}${[}Fe/H{]} & $\overline{\Delta}${[}$\alpha$/Fe{]} & $\overline{\Delta}$$T_\textrm{eff}$ & $\overline{\Delta}$log$g$ & $\overline{\Delta}${[}Fe/H{]} & $\overline{\Delta}${[}$\alpha$/Fe{]} \\ \midrule
\texttt{StarNet-INTRIGOSS} & 79 & 0.12 & 0.05 & 0.07 & 128 & 0.62 & 0.08 & 0.17 & - & - & - & - \\
\texttt{StarNet-FERRE} & 64 & 0.24 & 0.18 & 0.05 & 70 & 0.23 & 0.19 & 0.11 & 63 & 0.26 & 0.15 & 0.37 \\
\texttt{StarNet-MPIA} & 83 & 0.09 & 0.11 & 0.04 & 82 & 0.11 & 0.15 & 0.09 & 61 & 0.23 & 0.10 & 0.18 \\ \bottomrule
\end{tabular}
\end{table*}

\begin{figure*}
	\includegraphics[width=2.0\columnwidth]{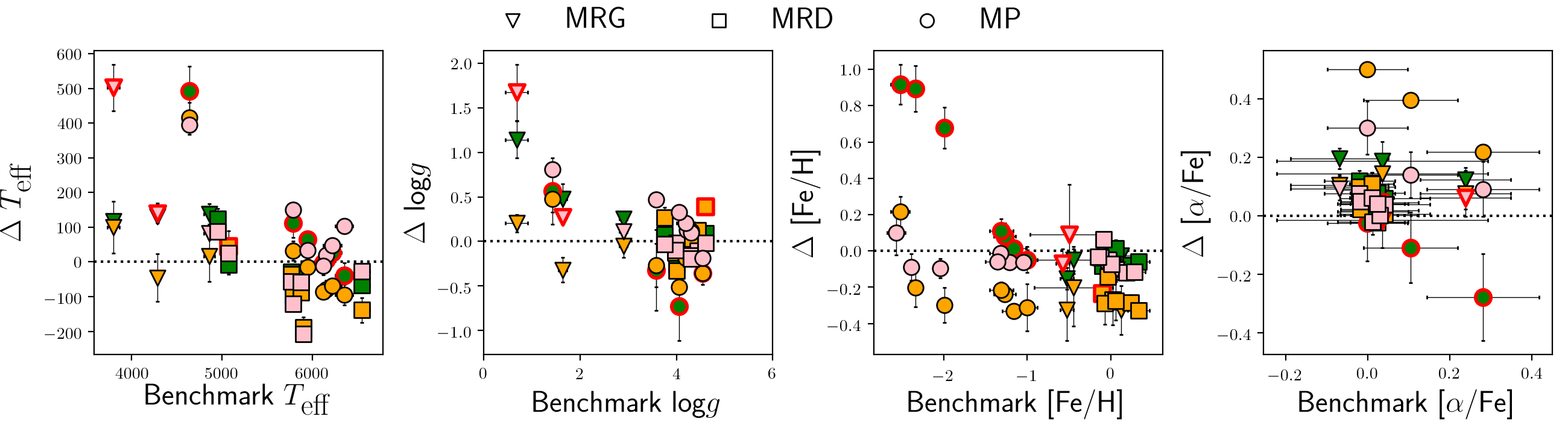}
    \caption{\StarNet was trained on the INTRIGOSS (green), FERRE (orange), and MPIA (pink) spectral grids, and each model was used to predict stellar parameters for the Gaia-ESO benchmark stars. The residuals between predictions and published values are shown here. The stars were split into metal-poor (MP) stars, metal-rich giants (MRGs) and metal-rich dwarfs (MRDs), following the procedure in R. Smiljanic et al. (2014). The red outlines indicate the stars lay outside any of the parameter ranges of the respective spectral grid. See Table \ref{tab:threemodels_benchmarks} for quantitative metrics. 
    \label{fig:benchmarks_uves_test} }
\end{figure*}

Following the procedure in \cite{smiljanic2014gaia}, the benchmark stars were separated into three groups in order to assess the \StarNet prediction accuracies in different regions of parameter space:

\begin{enumerate}[labelindent=8pt,labelwidth=\widthof{},itemindent=0em,leftmargin=!]
\item Metal-rich dwarf (MRD): [Fe/H] $>$ -1.00 and log$g$ $>$ 3.5
\item Metal-rich giant (MRG): [Fe/H] $>$ -1.00 and log$g$ $\leq$ 3.5
\item Metal-poor (MP): [Fe/H] $\leq$ -1.00
\end{enumerate}

The \StarNet training models are applied to the set of GES benchmark stars, including seven MRDs, three MRGs, and seven MP stars. The predictions are shown in Figure \ref{fig:benchmarks_uves_test}, plotted as the difference between the \StarNet model results and the GES benchmark parameter values.   All three versions of \StarNet provide reasonable estimates for the stellar labels of the benchmark stars, when those stars lay within the parameter range of the training sets.  Only one star stands out in the temperature predictions, HD~122563; the reason for this is not clear from our analysis, but we notice that this is true in all three models.   The INTRIGOSS model also appears to deviate at the lowest gravities.   We also notice that \texttt{StarNet-FERRE} results in lower metallicities than expected, however this is likely due to neglected NLTE effects, which are included in the GES benchmark abundances and \texttt{StarNet-MPIA} predictions (and indirectly the \texttt{StarNet-INRIGOSS} results due to its fine-turning, see Section \ref{sec:intrigoss_vs_mpia}).
\citealt{kovalev2019non} also report offsets in metallicity from the metal-poor benchmark stars that may imply we now have improved NLTE corrections. 

Table \ref{tab:threemodels_benchmarks} summarizes our results on the benchmark stars, noting that the metric for evaluating performance is the average quadratic difference, $\overline{\Delta}$, between the predictions and benchmark values (to be consistent with the analysis of \citealt{smiljanic2014gaia}). While the average quadratic difference removes knowledge of a positive or negative bias, it is a reliable metric for the overall discrepancy in predictions.
 
\begin{figure*}
	\includegraphics[width=2.0\columnwidth]{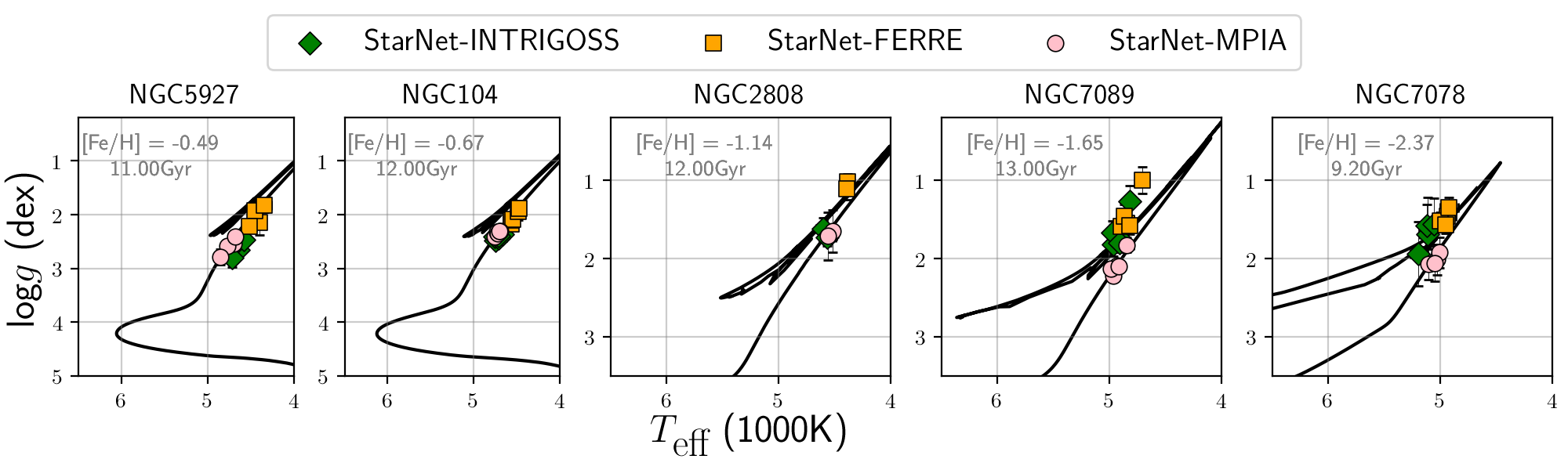}
    \caption{\texttt{StarNet} was separately trained on the INTRIGOSS (green), FERRE (orange), and MPIA (pink) synthetic spectral grids (\texttt{StarNet-INTRIGOSS}, \texttt{StarNet-FERRE}, \texttt{StarNet-MPIA}, respectively) and their predictions of $T_{\protect\rm eff}$ and log$g$ for a sample of the Gaia-ESO calibration cluster stars are compared with theoretical MIST isochrones \citep{choi2016mesa}. The isochrones were generated with ages and metallicities (shown in light grey text) extracted from the updated \protect\cite{harris2010new} catalog.}
    \label{fig:calibrating_clusters_isochrones}
\end{figure*}

\begin{figure*}
	\includegraphics[width=2.0\columnwidth]{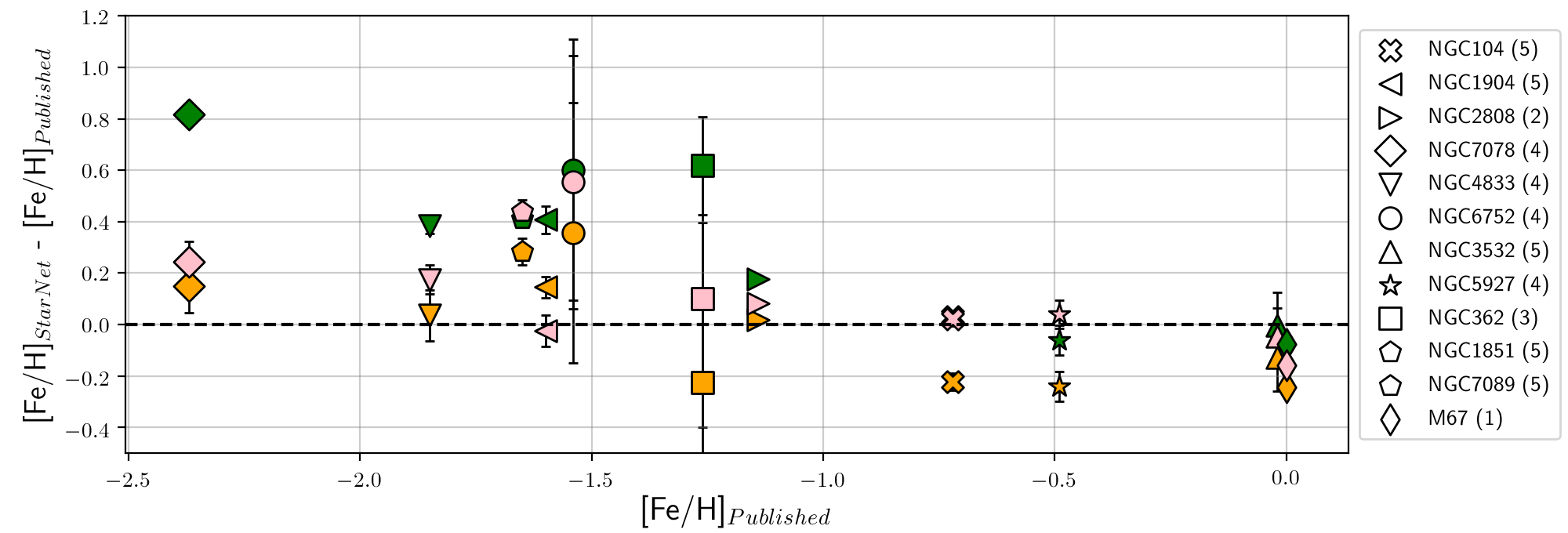}
    \caption{\StarNet was separately trained on augmented INTRIGOSS (green), FERRE (orange), and MPIA (pink) spectra, and shown here are predicted metallicities for a sample of calibration clusters from each model. The error bars indicate the standard deviation on the residual (except for M67, containing only one star, which shows the \StarNet uncertainty). Literature values were retrieved from the online updated catalog of \protect\cite{harris2010new} and the WEBDA database. Note the INTRIGOSS grid has a minimum metallicity of [Fe/H] = -1, so large discrepancies for metal-poor stars are expected. }
    \label{fig:calibrating_clusters_feh}
\end{figure*}

Altogether the results obtained through tests on the GES benchmark stars provide a convincing validation that our \StarNet application and training methods work well across a range of parameters \textit{for high S/N spectra}.  However, these benchmark stars are a statistically small sample, e.g., there are very few metal-poor giants.  Fortunately, the GES database also provides spectra and parameters for individual stars in several calibration clusters.  

\subsection{StarNet predictions for GES stars in clusters}

The FLAMES-UVES database includes spectra for individual stars in the globular clusters NGC7078, NGC104, NGC1851, NGC2808, NGC4833, NGC5927, NGC1904, and NGC6752, and the two open clusters M67 and NGC3532. 
Spectral determinations of $T_\textrm{eff}$ and log$g$ can also be compared to theoretical isochrones that are adjusted for distance and reddening.
For our \StarNet models, the predictions for individual stars in these globular clusters have been compared to the MESA Isochrones and Stellar Tracks (\textit{MIST}, \cite{choi2016mesa}), generated by adopting the metallicities and ages of each cluster from the Harris catalogue \citep{harris2010new}.  In Fig.~\ref{fig:calibrating_clusters_isochrones}, in general we find good overlap from all three \StarNet models -- over a range of metallicities -- with the isochrones, with overall better agreement from the \texttt{StarNet-MPIA} predictions. For some individual stars the \texttt{StarNet-FERRE} log$g$ values are more significantly offset from both the \texttt{StarNet-INTRIGOSS} and \texttt{StarNet-MPIA} results (e.g. NGC2808) and can also deviate from the isochrone positions (e.g. NGC7089). 

In Fig. \ref{fig:calibrating_clusters_feh}, the average metallicities from individual stars in each cluster are shown, with uncertainties derived from the standard deviation of the predictions.
We find good agreement with published data for the \texttt{StarNet-MPIA} and \texttt{StarNet-FERRE} models, which span the full metallicity range.  The \texttt{StarNet-INTRIGOSS} results deviate significantly for clusters below [Fe/H]\,=\,-1, as expected (this is outside the training parameter space).  
Again, we find that the \texttt{StarNet-MPIA} model provides $\sim 0.2$ dex better agreement than the \texttt{StarNet-FERRE} model, especially for clusters with [Fe/H]\,$>$\,-1.5.   

Finally, we note that the predictions from each of these \StarNet models are not calibrated.  Thus, the stellar parameters ($T_{\textrm{eff}}$, log$g$, and [Fe/H]) recovered by \StarNet are physically consistent for all stars in the training set (e.g., for both dwarfs and giants), at least to within the precision of the physics in the synthetic spectral grids.

\subsection{\StarNet predictions for the entire Gaia-ESO Survey (GES iDR4)}

\begin{figure*}
	\includegraphics[width=2.0\columnwidth]{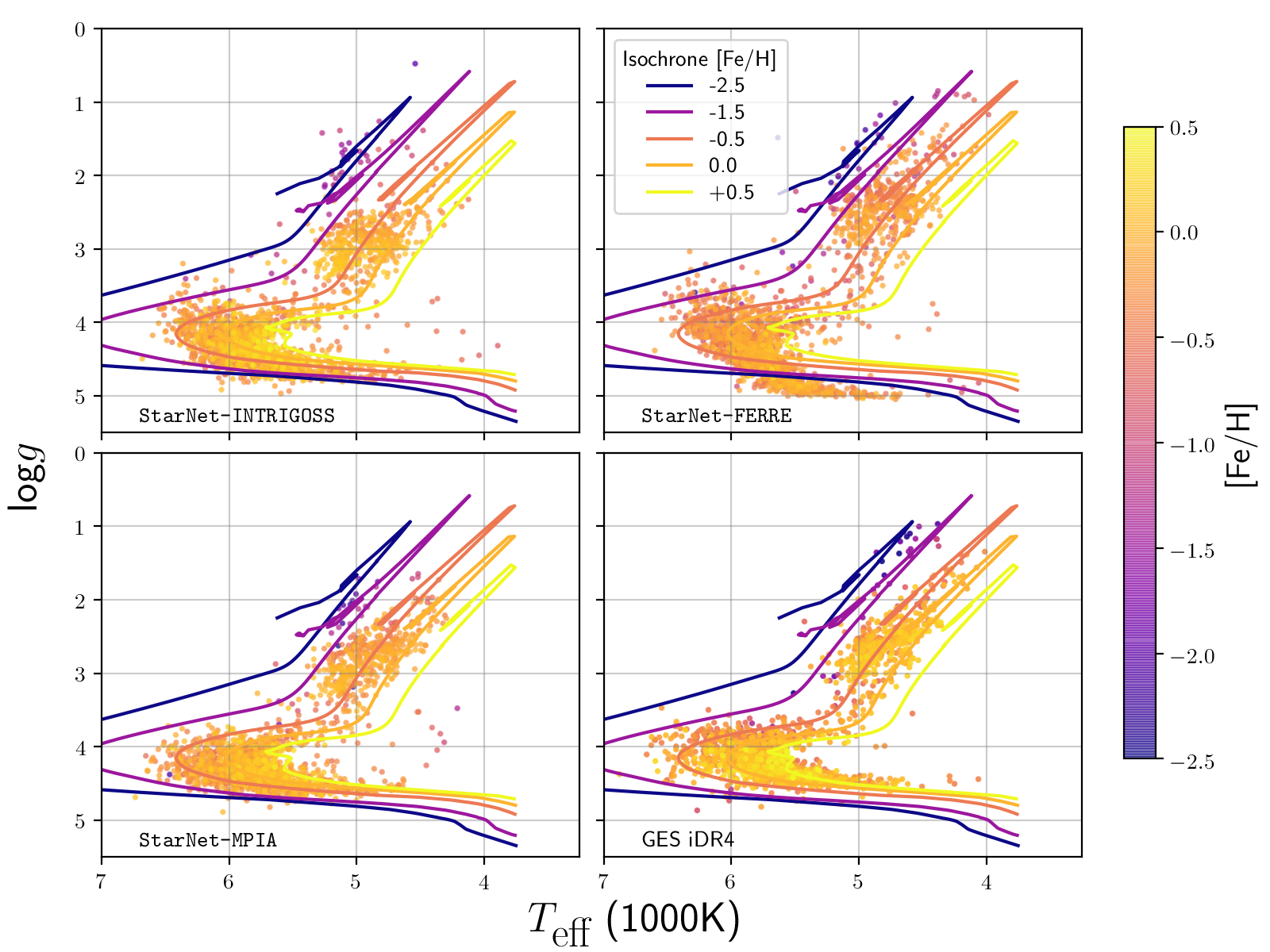}
    \caption{Kiel diagrams showing the physical consistency of \texttt{StarNet-INTRIGOSS}, \texttt{StarNet-FERRE}, and \texttt{StarNet-MPIA} predictions for $T_\textrm{eff}$, log$g$, and [Fe/H] on the test set of FLAMES-UVES spectra. Overlaying the predictions are MIST isochrones with an age of 8 Gyr and the metallicities shown. For comparison, the published GES iDR4 values are shown as well.}
    \label{fig:logg_teff_mpia_ferre_intri_gesidr4}
\end{figure*}

\begin{figure*}
	\includegraphics[width=2.0\columnwidth]{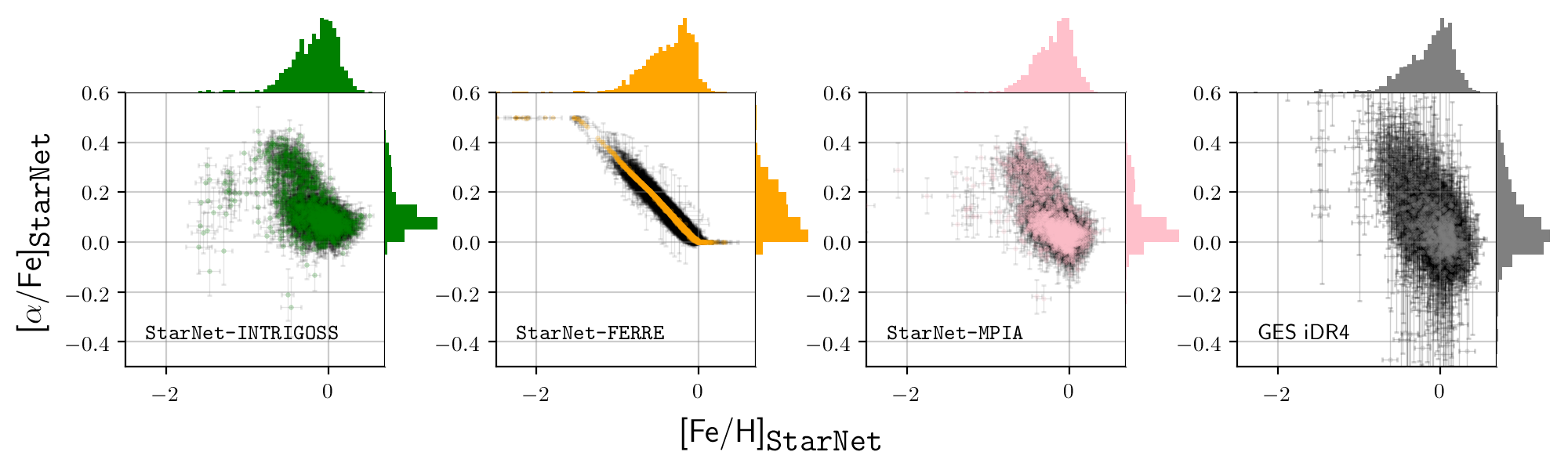}
    \caption{[$\alpha$/Fe] vs. [Fe/H] predictions of \texttt{StarNet-INTRIGOSS}, \texttt{StarNet-FERRE}, and \texttt{StarNet-MPIA} on the test set of FLAMES-UVES spectra. The GES iDR4 values are shown for comparison.  The predictions from \texttt{StarNet-FERRE} are poor because [$\alpha$/Fe] is a function of [Fe/H] in the FERRE grid adopted, whereas both \texttt{StarNet-INTRIGOSS} and \texttt{StarNet-MPIA} provide a much tighter distribution than seen from the GES iDR4 values.}
    \label{fig:alphafe_vs_feh}
\end{figure*}

 
The full catalogue of FLAMES-UVES spectra available in the GES database was examined with \texttt{StarNet}. Only a few selection cuts were made to produce a test sample from the observed spectra: stars were removed if (1) they had \texttt{NaN} values for any parameter in the GES iDR4 catalog, and (2) if the uncertainties produced by \StarNet for any parameter were abnormally large (our adopted limits were $\sigma T_\textrm{eff}$ \textgreater{} 65\,K,
$\sigma$[Fe/H] \textgreater{} 0.50,
$\sigma \textrm{log}g$ \textgreater{} 0.80,
$\sigma v_\textrm{rot}$ \textgreater{} 3\,\kms,
$\sigma v_\textrm{rad}$ \textgreater{} 5\,\kms).
These cuts decreased the full sample size from 2,308 individual stars to 2,200, rejecting a total of 108 stars (primarily those with very high radial velocities, $>$50\,km/s, beyond the training limits).

The $T_\textrm{eff}$, log$g$, and [Fe/H] predictions for the final sample are shown in Fig.~\ref{fig:logg_teff_mpia_ferre_intri_gesidr4} and compared to MIST isochrones.  We note that the GES iDR4 [Fe/H] values for this sample are 1DLTE results.
While predictions from \texttt{StarNet-FERRE} seem to fail for the dwarfs (possibly due to the coarse grid spacing), the log$g$ and $T_\textrm{eff}$ predictions for both \texttt{StarNet-INTRIGOSS} and \texttt{StarNet-MPIA} produce slightly higher values for the giants than GES iDR4 (yet still remain on the isochrones). The higher log$g$ and $T_\textrm{eff}$ values are pronounced for metal-poor stars, a trend that was also seen in \cite{kovalev2019non} due to the NLTE versus LTE metallicities.

The [$\alpha$/Fe] predictions are examined in Fig. \ref{fig:alphafe_vs_feh}, where the well-known pattern of a ``knee" occurs at a particular metallicity, presumably due to SN Ia contributions to iron at later times.  The knee is recovered for both the \texttt{StarNet-INTRIGOSS} and \texttt{StarNet-MPIA} models.  We also find that it is more tightly constrained in our models than the GES iDR4 values, implying that [$\alpha$/Fe] may be more precisely recovered from our supervised learning application.  The poor performance of \texttt{StarNet-FERRE} is expected, as [$\alpha$/Fe] is hardwired as a function of [Fe/H] in the FERRE grid we have adopted (i.e., it is not an independent grid dimension). 

\begin{figure*}
	\includegraphics[width=2.0\columnwidth]{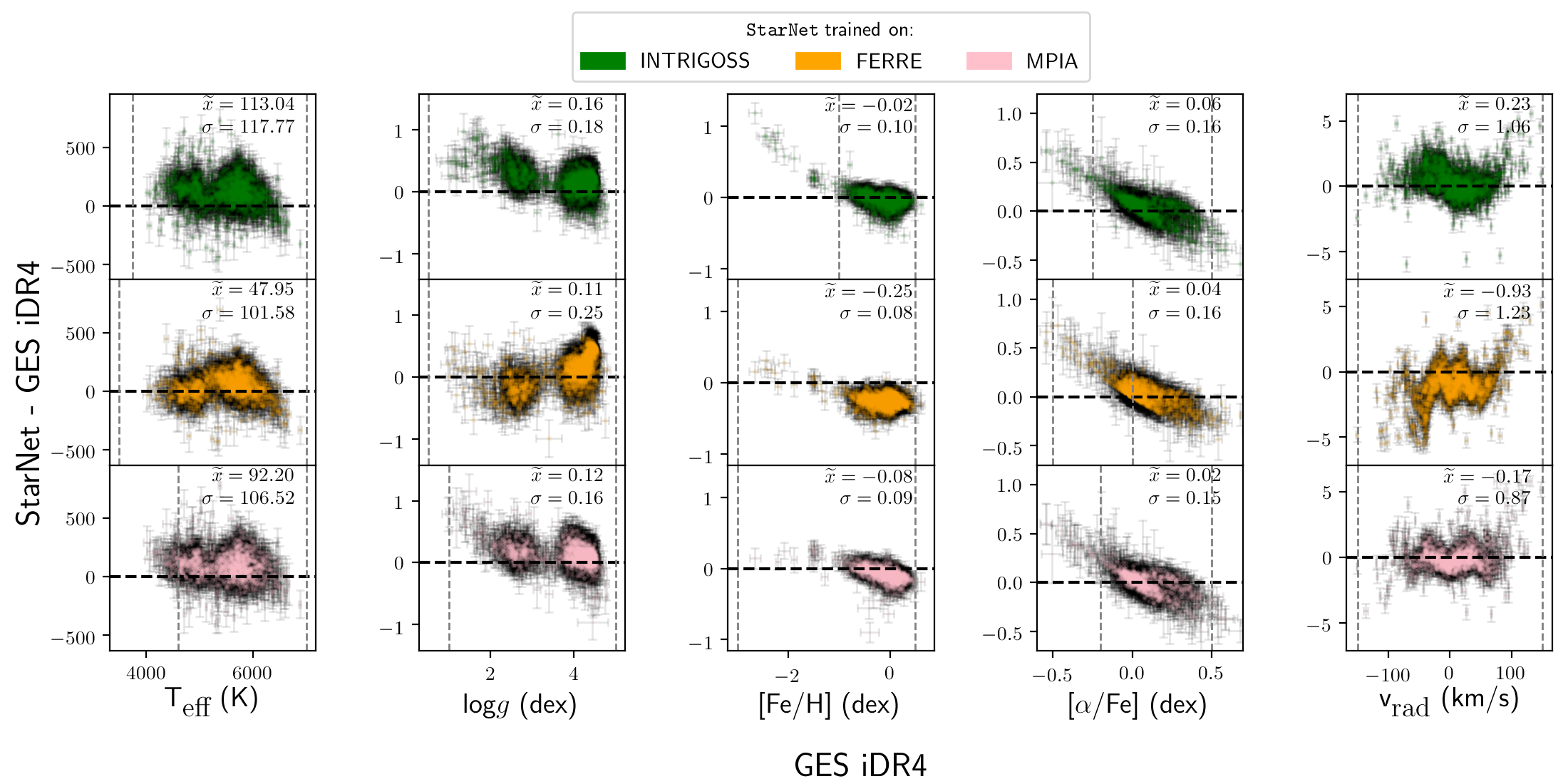}
    \caption{The stellar parameter predictions of \texttt{StarNet-INTRIGOSS} (green), \texttt{StarNet-FERRE} (orange), and \texttt{StarNet-MPIA} (pink) on 2,200 FLAMES-UVES spectra are compared to parameters from GES iDR4. The median ($\widetilde{x}$) and standard deviation ($\sigma$) of the residuals are shown as well. Vertical dashed lines correspond to the parameter ranges of the respective grid trained on.}
    \label{fig:uves_test}
\end{figure*}

In Fig.~\ref{fig:uves_test}, the residuals from all three \StarNet models are presented and compared to the GES iDR4 values for $T_{\rm eff}$, log$g$, [Fe/H], [$\alpha$/Fe], and $v_\textrm{rad}$.  We notice that the residuals in $T_{\rm eff}$ and log$g$ are slightly offset to larger values in \texttt{StarNet-MPIA} and \texttt{StarNet-INTRIGOSS} than for \texttt{StarNet-FERRE}.  Also, the metallicity residuals on the metal-poor stars from \texttt{StarNet-INTRIGOSS} are much larger than from the others since those stars are outside of its training parameter range.  The [$\alpha$/Fe] residuals from \texttt{StarNet-FERRE} are about the same size as from the other two, however the results are less reliable given that this is not an independent parameter in that grid.  And finally the $v_\textrm{rad}$ predictions from \texttt{StarNet-MPIA} are the most closely matched to GES-iDR4 values; however, all three models appear to predict values with $\ge$2x the observational errors ($\sim$0.4~km/s)\footnote{http://www.eso.org/rm/api/v1/public/releaseDescriptions/92}. The reason for this increased scatter of the radial velocities measured by \texttt{StarNet}, with respect to the values predicted with GES iDR4, is unclear. The increase is not seen while testing on (noisy) synthetic data (see Section \ref{section:methoddependentbiases} (v)). \StarNet robustness to wavelength calibration accuracy, and the translation equivariance properties of a CNN architecture, may influence the radial velocity precision. We defer this study for a future analysis.

\section{Discussion}

Our CNN spectral parameter application, \texttt{StarNet}, was originally developed and tested using the SDSS APOGEE near-IR observed and synthetic spectra databases \citep{fabbro2018}. In this paper, we have further developed \StarNet to include the prediction of uncertainties and the ability to train with any  collection of synthetic spectra (after augmentation).  Stellar parameter results are presented when \StarNet is trained with several different synthetic spectral grids, and tested on the optical FLAMES-UVES spectra from the Gaia-ESO Survey. This paper presents the first application of \StarNet to  optical spectral analyses, and provides a guideline on how to use synthetic spectra when training a neural network.

\subsection{Caveats for machine learning applications and the benefits of training on synthetic spectra}

The architecture and uncertainty methods for this iteration of \StarNet have been kept simple-- but precise and efficient.
This has been done on purpose to provide a recipe for any stellar spectroscopic survey.  We have shown that the choice of the synthetic grid may influence the stellar parameters accuracy more than any other source of uncertainty.

\subsubsection{Synthetic spectra are not a perfect training set}

Despite the encouraging results presented in this paper, it should be noted that training a neural network on synthetic spectra does pose problems; the physics of the stellar interiors that feeds into the synthesis of the model atmospheres, and the atomic physics required for precision radiative transfer calculations, is incomplete \citep[see e.g.,][]{amarsi2016, lind2017, barklem2016accurate}. Assumptions about the physics of stellar models (interiors and atmospheres) will affect the precision of any synthetic grid. This is especially true for certain types of stars, e.g. cool dwarfs where the formation mechanism for thousands of spectral absorption features remains unknown \citep[e.g.,][]{peterson2017, farbod2020}. 
Furthermore, modeling the effects of instrumental signatures and noise is not perfect, and reddening from interstellar dust is not accounted for when comparing observations to synthetic spectra.  Any mismatch between synthetic and observed spectra produces a synthetic gap which runs the risk of poor predictive power from \textit{any} prediction pipeline. 

Ongoing work aims to improve our understanding and implementation of stellar spectral physics.  As discussed previously, some groups are working to improve the theoretical basis for NLTE corrections in the formation of spectral features, while others are also exploring 3D modelling and other neglected or poorly constrained opacity effects.
The ``\textit{Including All the Lines}" project \citep{kurucz2011including} aims to compute better opacities in model atmospheres via a brute force approach of computing an ever increasing number of atomic and molecular line data. 
Machine learning approaches are also being examined for identifying unknown features or filling in gaps in unknown physics, e.g., through domain adaptation between synthetic and observed spectra \citep{obriain2020}. 
These efforts will help produce more accurate stellar parameters and chemical abundances for a larger variety and number of stars, in a consistent manner.

In cases where the synthetic spectra are not modelled correctly, there are various strategies to mitigate the errors when predicting on observed spectra. One example is to mask the parts of the spectra that are known to be modeled poorly \citep{ting2019}. Of course, it might be beneficial to skip training on synthetic spectra entirely, but training would then require a set of observed spectra which have accurate stellar parameters pre-determined through other methods (physical, \textit{non-spectroscopic}, to avoid implicit bias).  This is difficult for a large number of stars over a wide range in parameter space.  

\subsubsection{To train on synthetic or observed data?}

Training on a grid of synthetic spectra has the added benefit of not adopting correlations between stellar parameters which exist in the observed data. For example, when the bulk of a training set of observed spectra has a Mg-Al correlation, then a data-driven NN is more likely to falsely assign a Mg-Al correlation to globular cluster stars even if they are known {\it a priori} to be anti-correlated \citep[e.g., see the discussion by][]{leung2019}. This problem can be mitigated with domain knowledge, e.g. by windowing or weighting the spectra according to spectral features from a particular element. With synthetic spectra, an array of uncorrelated chemical abundances can be included in the synthesis of the spectral grids, though this could potentially lead to generating a prohibitive number of spectra.

A training set composed of observed spectra needs extra care to properly balance the dataset to cover uniformly the parameter space one wishes to predict from. Rare stars (e.g. carbon-enhanced metal-poor stars, ultra metal-poor stars, stars captured from nearby dwarf satellites, or r-process rich stars (see \citealt{Venn2020, Monty2020, Arentsen2019, Sakari2018, Kielty2017}), and even spectroscopic binaries \citep{merle2017gaia, merle2020gaia, el2018discovery, el2018signatures}, would be under-represented. If a training set does not include a significant proportion of peculiar stars, then predictions on these rare populations will lead to biased predictions. Data augmentation techniques can mitigate the bias; however, augmenting rare stars and binary spectra is not trivial. In machine learning applications, the training set is often the limiting factor, so special care is required to account for out-of-distribution samples. For data-driven methods, this problem is also difficult to address due to the smaller sample sizes. For synthetic grids, spectra of rare stars can be added on-demand.

In cases where the sample size of a spectroscopic survey is low (in the hundreds or low thousands of spectra), it \textit{might} be infeasible to train NN which produces accurate results within a supervised learning approach. This problem may also be overcome by synthetic spectra. The limits to the size of a synthetic training set are constrained primarily by the computing time required to produce the spectra. 

Another advantage to training with synthetic spectra is that a complete model and analysis pipeline can be created before first light is collected at the telescope.  Thus, as spectra are collected, a data reduction pipeline can reduce the data and also provide the stellar parameters, along with uncertainties in real time. Not only would this be a benefit to any science case, but it also permits for a real-time assessment of the spectral quality and accuracy of predictions.  This would provide valuable feedback necessary for queue observing and spectroscopic surveys.   

Overall, there are many benefits to using a neural network trained on synthetic spectra, though caution is necessary in selecting the synthetic grid.

\subsection{Comparing the various synthetic grids}

Five synthetic spectral grids have been examined in this paper. A comparison of the training and testing of \StarNet using the MPIA, INTRIGOSS, and FERRE synthetic grids was described in Section 4.  All discussions of the AMBRE and PHOENIX grids are provided in Appendix B. 

\subsubsection{INTRIGOSS and 1DNLTE/MPIA}
\label{sec:intrigoss_vs_mpia}
The line list used to generate the INTRIGOSS spectra was based on a semi-empirical calibration of standard stars, but without a physical underpinning.   
As described by \cite{franchini2018gaia}, the INTRIGOSS spectra were computed with atomic and molecular line lists \textit{modified} by tuning the oscillator strengths to reproduce a set of high-resolution reference spectra, namely the Solar spectrum and the GES spectra of five cool giants with high SNR ($>$100).
This makes sense when it is known that there are missing opacities and simplified assumptions in the 1DLTE radiative transfer and model atmosphere codes.  On the other hand, the MPIA synthetic grid includes NLTE corrections for several key elements with opacities and absorption lines in the optical spectra.
The similarities in metallicities between the INTRIGOSS and MPIA trained \StarNet models, for spectra with [Fe/H]$>-1.0$ (e.g., Figure \ref{fig:calibrating_clusters_feh}), suggests that the semi-empirical ``corrections" made to the INTRIGOSS line list can be (partially) explained as the missing NLTE corrections for some of the dominant opacity sources. Both are attempts to produce more realistic stellar spectra, but whereas the latter are motivated by a more complete understanding of radiative transfer in a stellar atmospheres, the former are made ad hoc to simply better represent calibration star spectra.  
This further suggests that the MPIA grid is physically the most suitable for scientific purposes and machine learning applications.

\subsubsection{Recommendations for Applications}

The success of the MPIA spectral training set in reproducing the GES iDR4 stellar parameters for the benchmark stars, globular clusters, and other survey stars, and with small residuals, leads us to recommend the \StarNet-MPIA model. Furthermore, the MPIA online synthetic generator permits individual abundances, $v_{\textrm{mic}}$, $v_{\textrm{mac}}$, and $v$sin$i$, to be varied, making it a powerful tool compared to static grids.  This should make it possible to test predictions for elemental abundances; however, some caution is needed since systematic errors may occur when a synthetic spectrum is generated with different chemistry from that adopted to build the model atmosphere  \citep{ting2016accelerated}.


Stellar parameters can be sufficiently well determined with 1DLTE models depending on the application and computational constraints. Indeed, 1DLTE grids still have a role in comparing with existing published catalogues and colour-temperature relationships, and were used recently in forecasts for chemical abundance precisions from various facilities, spectrograph resolutions, and wavelength ranges by \citet{sandford2020}. However, the current results show when more accuracy and realism are required, NLTE grids provide significant improvements
over LTE grids and should therefore play a prominent role in future studies.

\StarNet can also be trained for the fast and homogeneous analysis of existing spectral archives, such as the CFHT ESPaDOnS \citep{Donati2006} database, Gemini GRACES \citep{Chene2014} database, and upcoming Gemini GHOST spectrograph \citep[][to be commissioned by the end of  2020]{pazder2016}.   
The flexibility of these synthetic grids also means that \StarNet can be trained for lower resolution spectral archives as well, e.g., the SDSS BOSS database \citep{Dawson2016} or ESO X-SHOOTER library \citep{Vernet2011}.
Unfortunately, the current \StarNet setup requires retraining for each new observational data set, and/or for each new synthetic grid library.  In the future, this could be accelerated by using transfer learning techniques, e.g., training a very large neural network that would cover most cases and could be tuned to specific data sets or spectral parameters.

\subsection{Predicting chemical abundances from synthetic spectra}

To extend this analysis to predictions of chemical abundances, spectra could be produced within the parameter range of an existing grid, but not aligned with the grid points \citep[see][]{ting2019}. Indeed, producing spectra in a grid is inefficient within a high dimensional parameter space, as there will inevitably be multiple realizations of the same stellar parameter, resulting in an over abundance of spectra needed for a neural network analysis.   
It is more economical to produce spectra with randomly varying parameters  \citep[see][]{bergstra2012random}, especially when considering extending grids to \textgreater{}10 dimensions. This is the strategy that \cite{ting2019} adopted in generating a coarse sample of spectra to train on.
Sampling strategies for efficiently training deep networks is an active area of research which will naturally benefit the approach taken with our analysis.



\section{Conclusions}
\label{section:conclusions}


In this paper, we have presented an updated version of the \StarNet convolutional neural network used to calculate stellar parameters ($T_{\rm eff}$, log$g$, [Fe/H], [$\alpha$/Fe], $v_{\rm rot}$, and $v_{\rm rad}$) with good precision from high-resolution stellar spectra.  The main update to the neural network has been the implementation of deep ensembling to estimate realistic uncertainties in the predicted stellar parameters.   

\StarNet has been trained and tested with five independent grids of synthetic spectra (INTRIGOSS, FERRE, AMBRE, PHOENIX, and MPIA), highlighting its versatility.  
We use these grids to test our preferred \texttt{StarNet-MPIA} model, and estimate systematic offsets and uncertainties between the different spectral grids.    
We also show that data augmentation in the training sets can overcome the synthetic gap(s), which 
includes resolution matching, wavelength sampling, Gaussian noise and random zero flux values, applying rotational and radial velocities, continuum removal, and masking telluric regions.
Augmenting the training data with noise {\it before} the asymmetric sigma-clipping continuum estimation step is necessary to decrease the biases in predictions.

Once trained, each \StarNet model was able to predict the stellar parameters for  $\sim$2,300 FLAMES-UVES optical spectra for benchmark stars, individual stars in globular clusters, and other survey stars from the GES.  
The predictions from the \texttt{StarNet-MPIA} model, using NLTE spectra generated from an online tool \citep[see footnote 2,][]{kovalev2019non}, resulted in stellar parameters that (typically) had the smallest residuals when compared with the GES-iDR4 catalogue.  This is the only 1DNLTE synthetic grid tested here, although we note that the specifically-tuned 1DLTE INTRIGOSS grid also provides very good results within its limited parameter range.  We propose the ad hoc corrections made to the INTRIGOSS line list may (partially) mimic NLTE corrections derived from first principles.  The predictions and residuals for [$\alpha$/Fe] from the \texttt{StarNet-MPIA} model appear to be better constrained than the GES-iDR4 results.
 
We plan to train \StarNet for the analysis of optical spectra from Canadian observational facilities (CFHT ESPaDOnS, Gemini GRACES and GHOST), and to prepare for observational data from upcoming spectroscopic surveys, in a forthcoming publication. 
We are also developing new tools for more chemical abundance calculations with \texttt{StarNet}.  
Our codes are publicly available and simple to adapt to any set of synthetic spectra.



\section*{Acknowledgements}

We thank Jonay Gonz\`alez-Hern\`andez, David Aguado, Patrick de Laverny, 
Alejandra Recio-Blanco,
Szabolcs M\'esz\`aros, Mikhail Kovalev, and Maria Bergemann 
for many helpful discussions and access to their synthetic spectral grids. We are grateful to Balaji Lakshminarayanan for helpful feedback in using the deep ensembling method, and Henry Leung for his work and discussions in improving \StarNet.
We also thank the anonymous referee for directing us to the 1DNLTE synthetic spectrum generator at MPIA and other helpful comments that improved this paper. SB, SF, and KAV thank the Natural Sciences and Engineering Research Council for funding through the Discovery Grants program and the CREATE program in New Technologies for Canadian Observatories.  NK thanks Mitacs for funding through their 2019 Globalink Research Internship program.

\section*{Data availability}

The raw MPIA spectra generated for this work are publicly available (see footnote 3). All other data underlying this article will be shared on reasonable request to the corresponding author. 



\bibliographystyle{mnras}
\bibliography{bibliography}



\newpage
\appendix

\section{Continuum estimation}
\label{appendix:continuum}

Special attention is required for good estimates of the stellar continuum in a  spectroscopic analysis. Any method used for estimating the continuum should be invariant to both the shape and the signal-to-noise (S/N) of the spectrum to prevent the introduction of noise-dependent biases into the parameter estimations.

Several existing methods for continuum estimation involve polynomial fits, with some research groups selecting high order polynomial fits to the entire spectrum, and others fitting a lower order polynomial to a set of identified `continuum pixels' \citep{Casey2016}. Other popular methods involve splitting the spectrum into short segments of equal length and estimating the continuum of each segment \citep[e.g.,][]{ASPCAP2016, Ness2015}. The segment methods perform well in cases where the spectral shape varies significantly over the wavelength range, possibly due to different detectors.

\begin{figure*}
	\includegraphics[width=2.0\columnwidth]{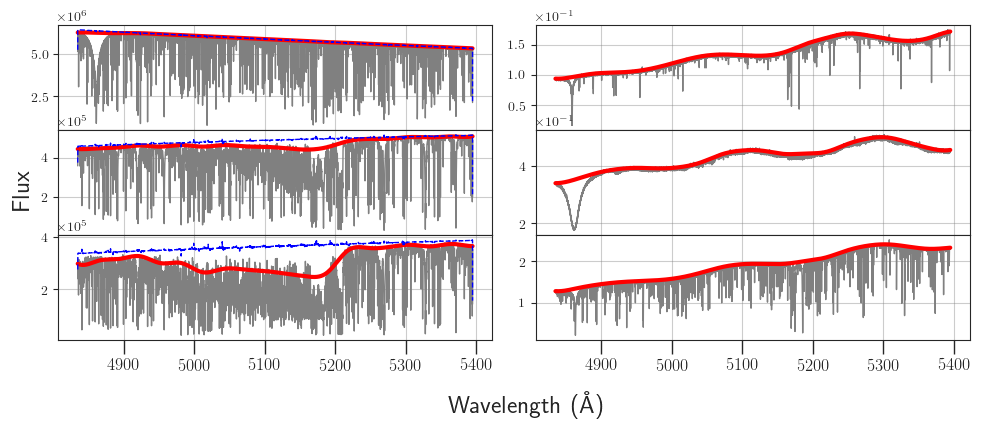}
    \caption{The results of our continuum fitting procedure for a random sample of INTRIGOSS synthetic spectra (left column) and FLAMES-UVES spectra (right column). The red line indicates the estimated continuum, and for the INTRIGOSS spectra the blue dashed line indicates the true continuum. The complex cyclical shape of the FLAMES-UVES spectra eludes simple fits of polynomials.
    \label{fig:continuum_estimates}}
\end{figure*}

In this paper, a method based on segmenting the spectra was adopted: with each segment of 10 Angstroms, the known strong absorption features are masked, then iteratively the median of the segment flux values is found and flux values are rejected above and below when discrepant by 2 and 0.5 standard deviations, respectively, until convergence is achieved. This `asymmetric sigma clipping' more aggressively rejects absorption features in order to find the true continuum. Once the continuum has been estimated in each segment, a cubic spline is fit to the segments. Figure \ref{fig:continuum_estimates} shows the ability of this method to fit both the complex shape of VLT/UVES spectra and the synthetic INTRIGOSS spectra.

\begin{figure}
	\includegraphics[width=1.0\columnwidth]{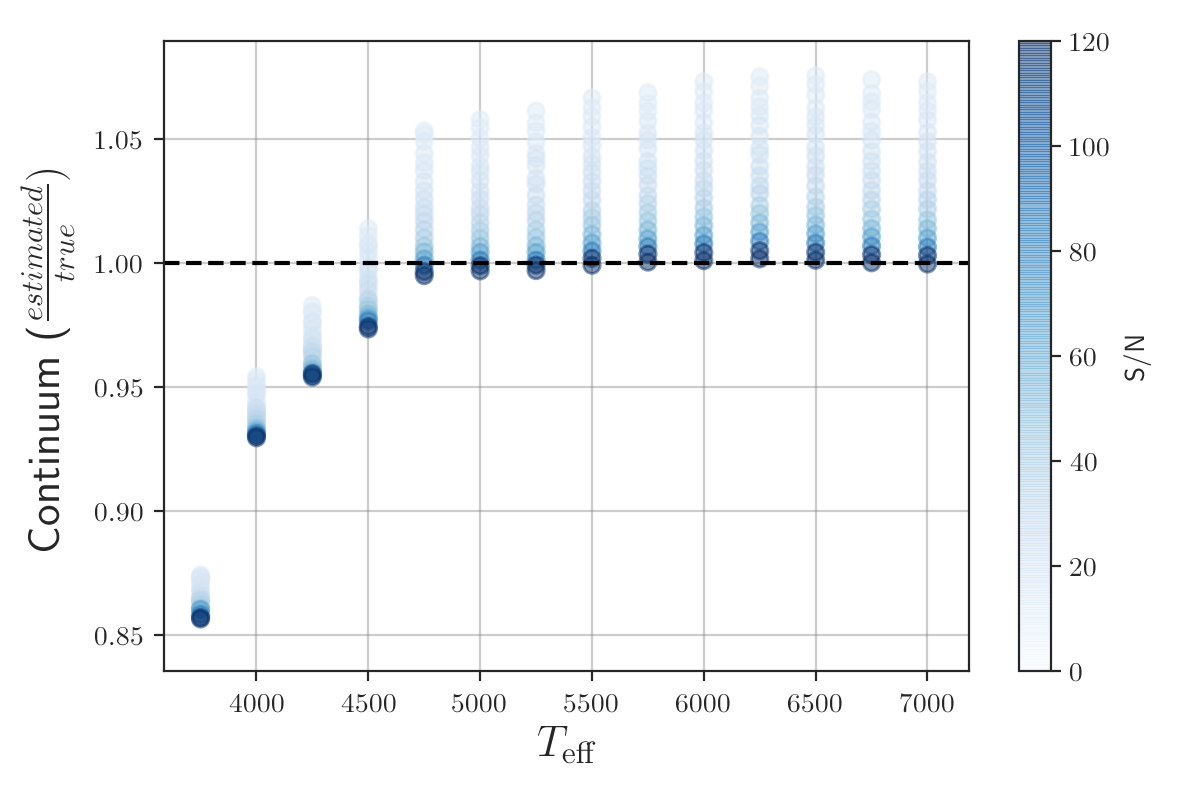}
    \caption{The systematic bias in the asymmetric sigma clipping method for the continuum estimation. Each INTRIGOSS spectrum was modified by varying the Gaussian noise, estimating the continuum, and averaging the offset from the true continuum.  The median offsets shown here for all INTRIGOSS spectra were derived in bins of noise and temperature. At the lowest temperatures, most of the spectrum lies below the true continuum due strong absorption features. }
    \label{fig:noise_dependent_bias}
\end{figure}

A known caveat with the asymmetric sigma clipping method is its noise dependent bias: as the noise levels increase in a spectrum, the found continuum is pushed further towards the `noise ceiling', and thus the estimated continuum is above the true continuum. 
Figure~\ref{fig:noise_dependent_bias} shows this bias as a function of temperature.  It can be seen that in all cases the estimated continuum for a set of synthetic spectra, where the true continuum is known a priori, is higher (by up to several percent) for a noisy spectrum. Also shown is the trend of spectra with lower temperatures to have a continuum estimate well below the true continuum.  This is expected since the majority of a low temperature spectrum lies below the continuum (due to extensive line blanketing), but this is not a problem here since this trend exists in both the synthetic and observed spectra. If the estimated continuum is significantly higher than the true continuum, the resulting continuum-normalized spectra will contain artificially lowered flux values. This would lead to deeper absorption features which could mimic a lower temperature or higher metallicity than the true value. 

\begin{figure*}
	\includegraphics[width=2.0\columnwidth]{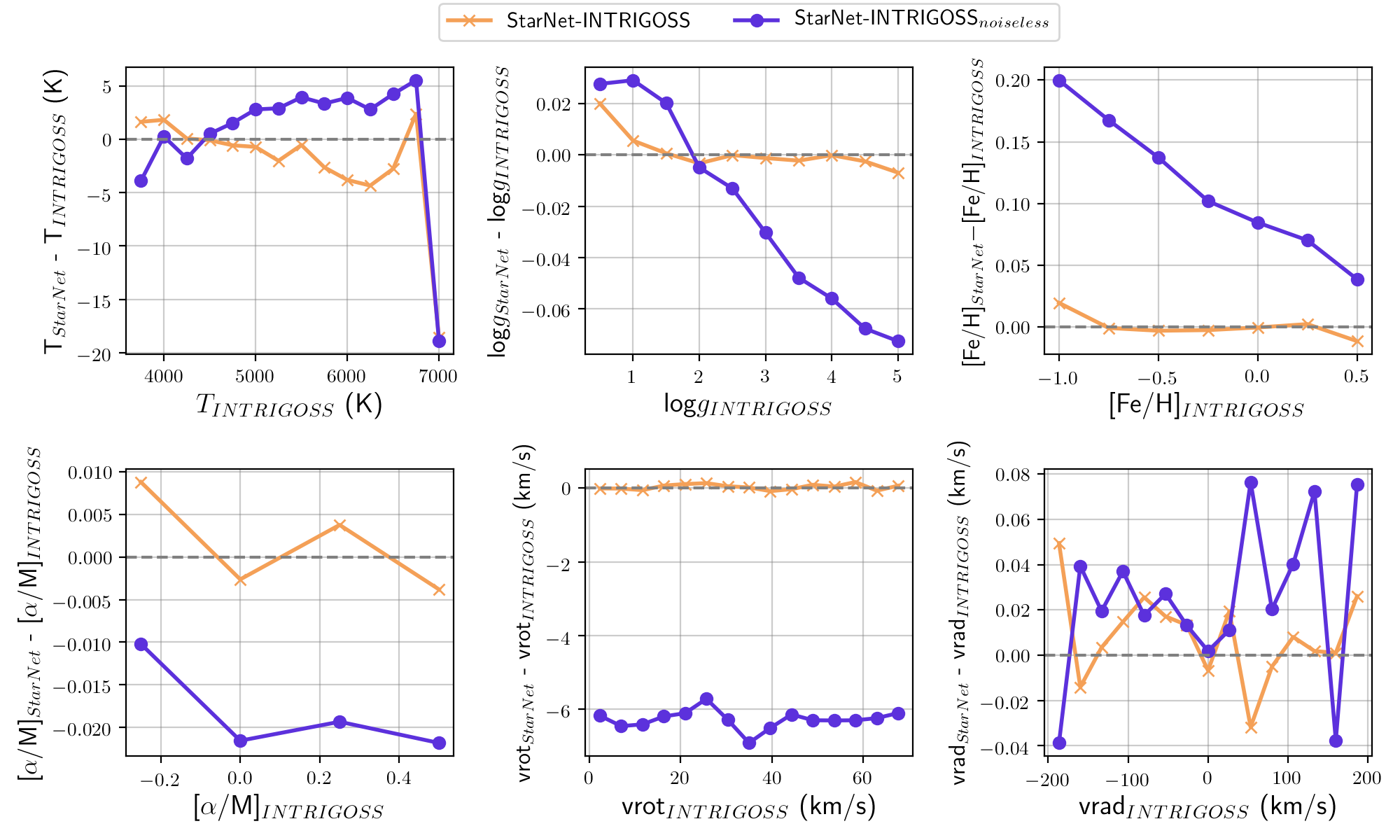}
    \caption{Residual plots to show  noise-dependent biases from the asymmetric sigma clipping continuum removal in the stellar parameter estimations.   
    Two versions of \StarNet\ were trained: one model, \texttt{StarNet-INTRIGOSS} (orange), was trained on 90,000 INTRIGOSS spectra augmented as outlined in Section \ref{section:augmenting}, and the other, \texttt{StarNet-INTRIGOSS}$_{\mathrm{noiseless}}$ (purple), was trained identically except without the addition of noise to the synthetic spectra prior to continuum removal. Each was tested on 10,000 noisy INTRIGOSS spectra, the median residual at each grid point was calculated, and the results for all spectra with S/N \textless{} 80 are shown here. The discrepancies are the most pronounced at lower metallicities, higher surface gravities, and across all rotational velocities.}
    \label{fig:starnet_INTRIGOSS_noise_test}
\end{figure*}

\begin{figure}
	\includegraphics[width=1.0\columnwidth]{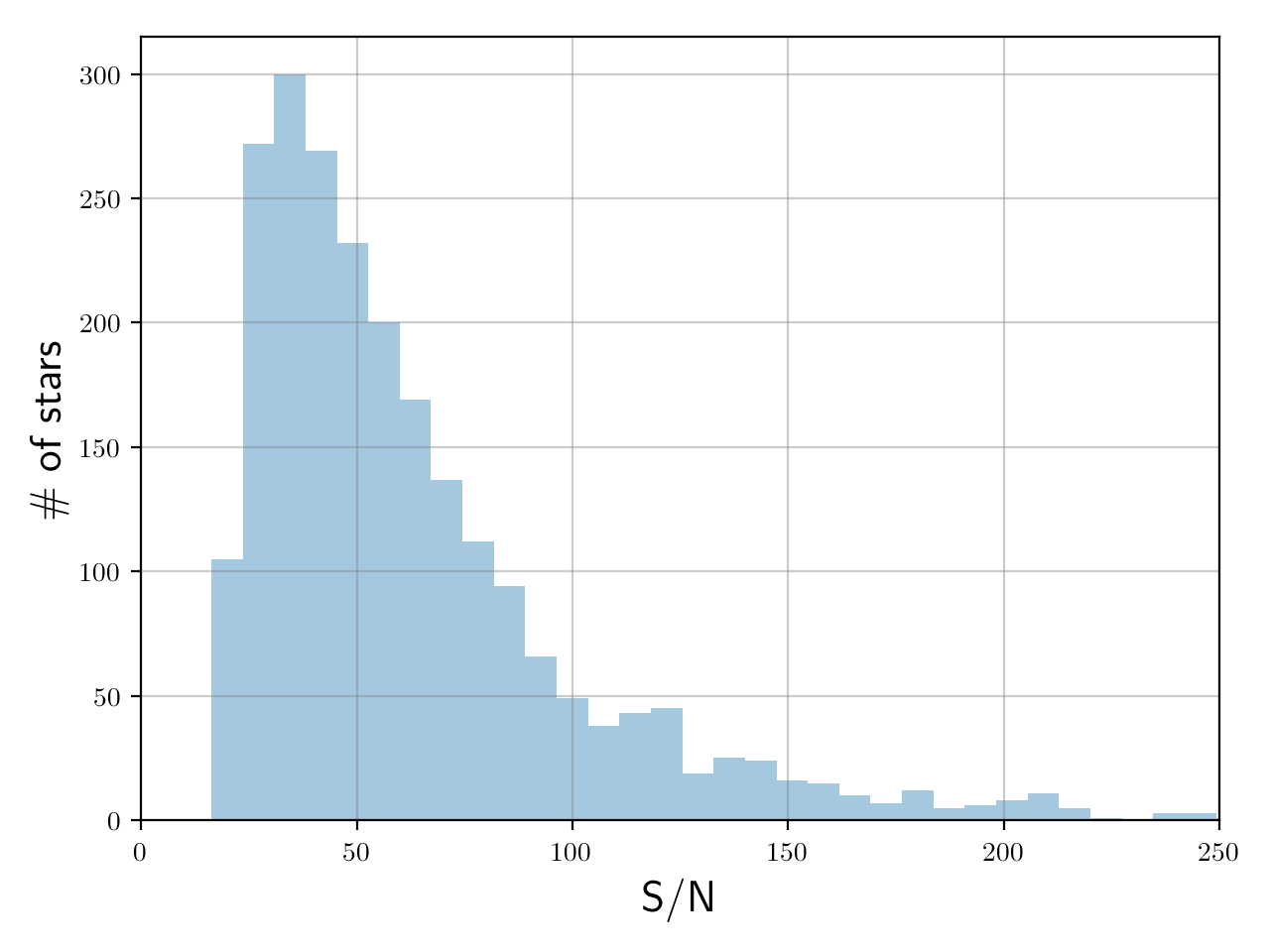}
    \caption{The S/N distribution of the Gaia-ESO FLAMES-UVES spectra.}
    \label{fig:snr_distribution_uves}
\end{figure}

To assess the impact of continuum fitting due to noise, \texttt{StarNet-INTRIGOSS} was trained with noiseless synthetic spectra and with Gaussian noise added (augmentation step (v) in Section~\ref{section:augmenting}).
Both of these trained models were tested on a set of 10,000 augmented (noisy) INTRIGOSS spectra, and the predictions for both models on all spectra with S/N \textless{} 100 are shown in Fig.~\ref{fig:starnet_INTRIGOSS_noise_test}
(the S/N distribution for the GES data is shown in Fig.~\ref{fig:snr_distribution_uves}). 
As expected, there are clear biases for all stellar parameters when \StarNet is trained on noiseless spectra, with more prominent discrepancies at low metallicities, high surface gravities, and across all rotational velocities. These biases are reduced when trained with noisy spectra; by adding noise to the spectra before the continuum removal step in the pre-processing stage, the neural network can learn to compensate for noise-dependent bias. 
Although this bias dependence is smooth, and it can be corrected in other ways and in other methodologies, the neural network compensates for it automatically.  Furthermore, the flexibility of the neural network means that it has the potential to handle even more complex bias dependencies (e.g.,  persistence in some of the early APOGEE spectra; see \citealt{Jahandar2017}).



 
Other continuum estimation techniques were examined, e.g. Gaussian smoothing normalization \citep{ho2017label}, but they were found to affect the synthetic spectra differently than the observed spectra and led to more discrepant results.

\newpage


\section{Results of training on the AMBRE and PHOENIX grids}

\begin{figure*}
	\includegraphics[width=2.0\columnwidth]{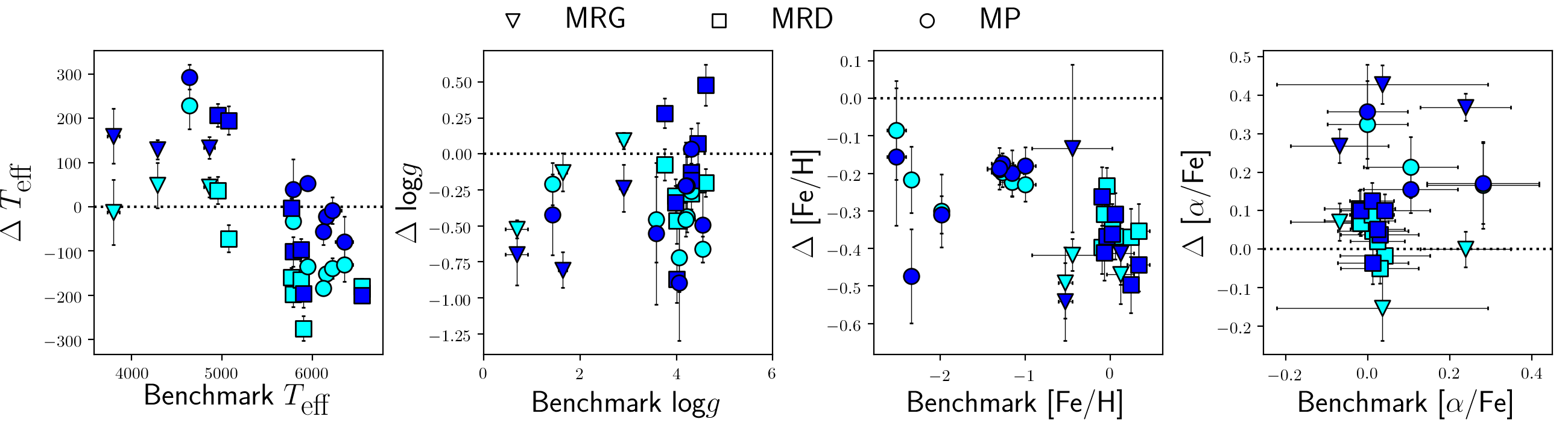}
    \caption{Similar to Figure~\ref{fig:benchmarks_uves_test} but here \texttt{StarNet} was instead trained on the AMBRE (cyan) and PHOENIX (blue) spectral grids to compare predicted stellar parameters for the Gaia-ESO benchmark stars. The residuals between predictions and published values are shown here. The stars were split into metal-poor (MP) stars, metal-rich giants (MRGs) and metal-rich dwarfs (MRDs), following the procedure in R. Smiljanic et al. (2014). See Table \ref{tab:ambrephoenix_benchmarks} for quantitative metrics. }
    \label{fig:benchmarks_phoenix_ambre}
\end{figure*}

\begin{figure*}
	\includegraphics[width=2.0\columnwidth]{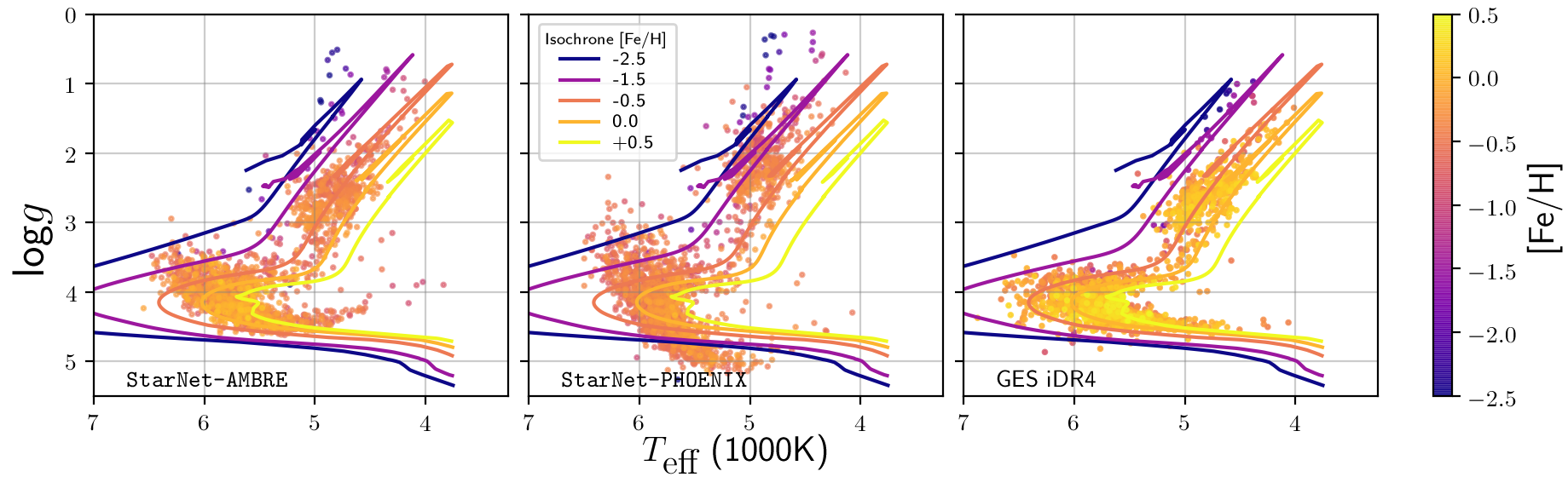}
    \caption{Similar to Figure~\ref{fig:logg_teff_mpia_ferre_intri_gesidr4}, these are Kiel diagrams showing the physical consistency of \texttt{StarNet-AMBRE} and \texttt{StarNet-PHOENIX} predictions for $T_\textrm{eff}$, log$g$, and [Fe/H] on the test set of FLAMES-UVES spectra. Overlaying the predictions are MIST isochrones with an age of 8 Gyr and the metallicities shown. For comparison, the published GES iDR4 values are shown as well.}
    \label{fig:kiel_phoenix_ambre}
\end{figure*}

\begin{figure*}
	\includegraphics[width=2.0\columnwidth]{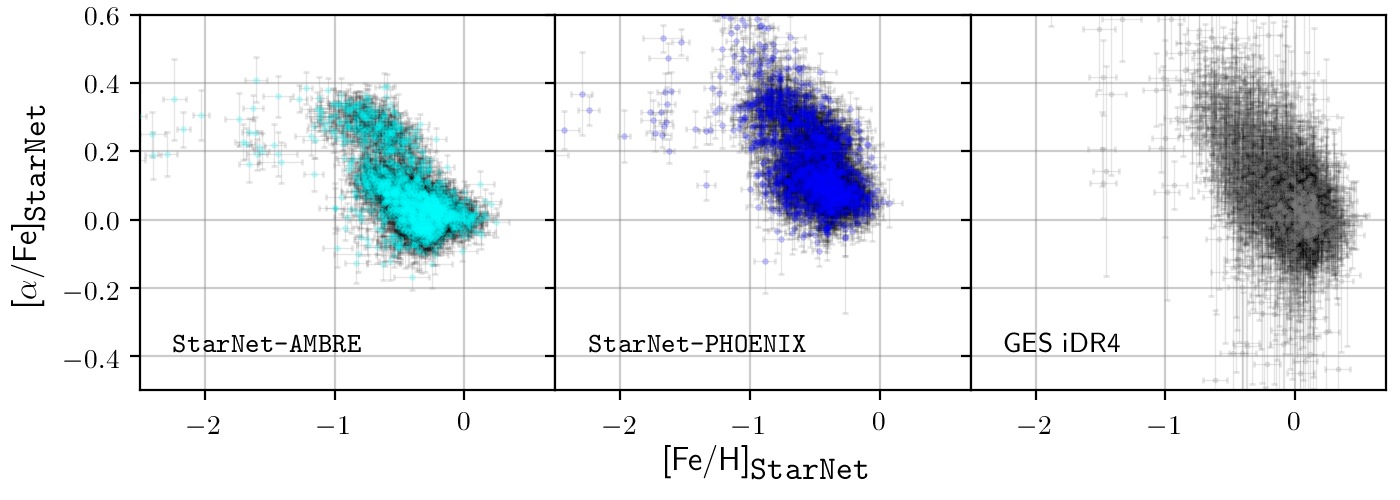}
    \caption{Similar to Figure~\ref{fig:alphafe_vs_feh}, [$\alpha$/Fe] vs. [Fe/H] predictions of \texttt{StarNet-AMBRE} and \texttt{StarNet-PHOENIX} on the test set of FLAMES-UVES spectra. Also plotted are the GES iDR4 values for comparison.}
    \label{fig:alphafe_vs_feh_phoenix_ambre}
\end{figure*}

An examination of the impact of training \StarNet with the AMBRE and PHOENIX spectral grids is provided in this Appendix.   In general, we found both sets of spectra provided worse results than the INTRIGOSS, FERRE, and our MPIA grids, when applied to the Gaia-ESO spectral database and compared to the GES iDR4 results.
In Fig. \ref{fig:benchmarks_phoenix_ambre} and Table \ref{tab:ambrephoenix_benchmarks}, it is clear that the benchmark stars residuals in all of the stellar parameters are larger than they were when trained on the other grids, especially for metallicity (with the exception of metal-poor stars with INTRIGOSS, which are beyond its training range).   In fact, these grids provide systematically lower metallicities at $\ge -0.1$ dex for all of the benchmark stars.  
This result is further emphasised in Fig. \ref{fig:kiel_phoenix_ambre}, where the dwarfs and subgiants are poorly fit and tending towards lower metallicities than the GES-iDR4 results.  Furthermore, in Fig. \ref{fig:alphafe_vs_feh_phoenix_ambre}, a slight offset towards larger [$\alpha$/Fe] values is also likely due to the slightly lower [Fe/H] results.
The cause of the poor predictions when \StarNet is trained on AMBRE or PHOENIX spectra is unknown, though outdated atomic data for PHOENIX grids is a potential source of discrepancy.

\begin{table*}
\caption{A comparison of stellar parameter results from \StarNet trained on the AMBRE and PHOENIX augmented grids and applied to GES benchmark stars. MRD = metal rich dwarfs, MRG = metal rich giants, and MP = metal poor stars.  The average quadratic differences (see text) between the StarNet predictions and the GES benchmark star parameters (for those stars only within the parameter ranges trained on) are shown.}
\label{tab:ambrephoenix_benchmarks}
\hspace*{-0.0cm}\begin{tabular}{@{}lcccccccccccc@{}}
\toprule
 & \multicolumn{4}{c}{MRD (7 stars)} & \multicolumn{4}{c}{MRG (3 stars)} & \multicolumn{4}{c}{MP (7 stars)} \\
 \cmidrule(lr){2-5}\cmidrule(lr){6-9}\cmidrule(lr){10-13}
 & $\overline{\Delta}$$T_\textrm{eff}$ & $\overline{\Delta}$log$g$ & $\overline{\Delta}${[}Fe/H{]} & $\overline{\Delta}${[}$\alpha$/Fe{]} & $\overline{\Delta}$$T_\textrm{eff}$ & $\overline{\Delta}$log$g$ & $\overline{\Delta}${[}Fe/H{]} & $\overline{\Delta}${[}$\alpha$/Fe{]} & $\overline{\Delta}$$T_\textrm{eff}$ & $\overline{\Delta}$log$g$ & $\overline{\Delta}${[}Fe/H{]} & $\overline{\Delta}${[}$\alpha$/Fe{]} \\ \midrule
\texttt{StarNet-AMBRE} & 155 & 0.25 & 0.34 & 0.05 & 47 & 0.11 & 0.48 & 0.04 & 129 & 0.50 & 0.23 & 0.24 \\
\texttt{StarNet-PHOENIX} & 134 & 0.31 & 0.40 & 0.09 & 131 & 0.52 & 0.48 & 0.32 & 43 & 0.40 & 0.25 & 0.26 \\ \bottomrule
\end{tabular}
\end{table*}

\bsp	
\label{lastpage}
\end{document}